\documentclass{aa}
\usepackage{graphicx}                    
\usepackage{color}                       
\usepackage{hyperref}
\usepackage{breakurl}                         

\usepackage{amsmath}
\usepackage{amsfonts} 

\usepackage{cool} 
\usepackage{breqn} 
\usepackage{bm} 
\usepackage{natbib}

\hypersetup{
  colorlinks=true,   
  citecolor=blue,
  urlcolor=blue,     
  linkcolor=red,     
}

\newcommand{\fref}[1]{Figure~\ref{#1}}
\newcommand{\eref}[1]{Equation~\eqref{#1}}
\newcommand{\sref}[1]{Section~\ref{#1}}
\newcommand{\axref}[1]{Appendix~\ref{#1}}
\newcommand{\tref}[1]{Table~\ref{#1}}

\newcommand{\abs}[1]{\ensuremath{\lvert{#1}\rvert}}

\newcommand{\pp}{\ensuremath{\delta\! p}}

\newcommand{\Xip}{\ensuremath{\delta \bm{\xi}}}
\newcommand{\Xirp}{\ensuremath{\delta \xi_r}}
\newcommand{\Xiperp}{\ensuremath{\delta \xi_{\perp}}}
\newcommand{\Xipara}{\ensuremath{\delta \xi_{\parallel}}}
\newcommand{\Xiphip}{\ensuremath{\delta \xi_{\varphi}}}
\newcommand{\Xizp}{\ensuremath{\delta \xi_z}}
\newcommand{\Xirip}{\ensuremath{\delta \xi_{r i}}}
\newcommand{\Xirep}{\ensuremath{\delta \xi_{r e}}}
\newcommand{\Bp}{\ensuremath{\delta\! \bm{B}}}
\newcommand{\Brp}{\ensuremath{\delta\! B_r}}
\newcommand{\Bphip}{\ensuremath{\delta\! B_{\varphi}}}
\newcommand{\Bphii}{\ensuremath{B_{\varphi i}}}
\newcommand{\Bphie}{\ensuremath{B_{\varphi e}}}
\newcommand{\Bzp}{\ensuremath{\delta\! B_z}}
\newcommand{\pTp}{\ensuremath{\delta\! p_T}}
\newcommand{\pTip}{\ensuremath{\delta\! p_{T i}}}
\newcommand{\pTep}{\ensuremath{\delta\! p_{T e}}}

\newcommand{\mee}{\kappa}

\newcommand{\Jeq}{\ensuremath{\bm{J}}}
\newcommand{\Beq}{\ensuremath{\bm{B}}}

\newcommand{\Beqz}{\ensuremath{B_z}}
\newcommand{\Beqphi}{\ensuremath{B_{\varphi}}}

\newcommand{\Beqzi}{\ensuremath{B_{z i}}}
\newcommand{\Beqphii}{\ensuremath{B_{\varphi i}}}

\newcommand{\Beqze}{\ensuremath{B_{z e}}}
\newcommand{\Beqphie}{\ensuremath{B_{\varphi e}}}
\newcommand{\peq}{\ensuremath{p}}
\newcommand{\peqi}{\ensuremath{p_{i}}}

\newcommand{\rhoeq}{\ensuremath{\rho}}
\newcommand{\rhoi}{\ensuremath{\rho_i}}
\newcommand{\rhoe}{\ensuremath{\rho_e}}

\newcommand{\kwvTh}{\ensuremath{k_{\varphi}}}
\newcommand{\kwvZ}{\ensuremath{k_{z}}}
\newcommand{\kwvRin}{\ensuremath{k_{r i}}}
\newcommand{\kwvRout}{\ensuremath{k_{r e}}}
\newcommand{\kwvRIN}{\ensuremath{k_{r I}}}
\newcommand{\kwvROUT}{\ensuremath{k_{r E}}}

\newcommand{\NI}{\ensuremath{N_I}}
\newcommand{\Ni}{\ensuremath{n_i}}
\newcommand{\NE}{\ensuremath{N_E}}
\newcommand{\Ne}{\ensuremath{n_e}}

\newcommand{\tuberadius}{\ensuremath{r_{a}}}
\newcommand{\kummerVar}{\ensuremath{s}}
\newcommand{\RO}{\ensuremath{\mathcal{K}}}
\newcommand{\NU}{\ensuremath{\nu}}
\newcommand{\omegaA}{\ensuremath{\omega_{A}}}
\newcommand{\omegaAi}{\ensuremath{\omega_{A i}}}
\newcommand{\omegaAe}{\ensuremath{\omega_{A e}}}
\newcommand{\omegaC}{\ensuremath{\omega_{c}}}
\newcommand{\omegaCi}{\ensuremath{\omega_{c i}}}
\newcommand{\omegaCe}{\ensuremath{\omega_{c e}}}
\newcommand{\vAphI}{\ensuremath{v_{A\Phi I}}}
\newcommand{\vAphE}{\ensuremath{v_{A\Phi E}}}
\newcommand{\vAphi}{\ensuremath{v_{A\varphi i}}}
\newcommand{\vAphe}{\ensuremath{v_{A\varphi e}}}
\newcommand{\vA}{\ensuremath{v_{A}}}
\newcommand{\vAI}{\ensuremath{v_{A I}}}
\newcommand{\vAi}{\ensuremath{v_{A i}}}
\newcommand{\vAE}{\ensuremath{v_{A E}}}
\newcommand{\vAe}{\ensuremath{v_{A e}}}
\newcommand{\vS}{\ensuremath{v_{s}}}
\newcommand{\vSI}{\ensuremath{v_{S I}}}
\newcommand{\vSi}{\ensuremath{v_{s i}}}
\newcommand{\vSE}{\ensuremath{v_{S E}}}
\newcommand{\vSe}{\ensuremath{v_{s e}}}
\newcommand{\vF}{\ensuremath{v_{F}}}
\newcommand{\vph}{\ensuremath{v_{ph}}}
\newcommand{\vTI}{\ensuremath{v_{T I}}}
\newcommand{\vTi}{\ensuremath{v_{T i}}}
\newcommand{\vTE}{\ensuremath{v_{T E}}}
\newcommand{\vTe}{\ensuremath{v_{T e}}}
\newcommand{\kummerA}{\ensuremath{a}}
\newcommand{\kummerB}{\ensuremath{b}}
\newcommand{\KummerM}[3]{\ensuremath{M\!\!\left(#1,#2; #3 \right)}}
\newcommand{\KummerU}[3]{\ensuremath{U\!\!\left(#1,#2; #3 \right)}}
\newcommand{\TwistConst}{\ensuremath{S}}
\newcommand{\TwistConstI}{\ensuremath{S_i}}
\newcommand{\TwistConstE}{\ensuremath{S_e}}

\newcommand{\twopartdef}[4]
{
	\left\{
		\begin{array}{ll}
			#1 & \mbox{for } #2 \\
			#3 & \mbox{for } #4
		\end{array}
	\right.
}

\begin{document}


\twocolumn[{%
\vspace*{4ex}
\begin{center}
  {\Large \bf Axisymmetric Modes in Magnetic Flux Tubes with Internal and External Magnetic Twist}\\[4ex]       
  {\large \bf I.~Giagkiozis$^{1}$, 
              V.~Fedun$^{1,2}$,
              R.~Erd\'{e}lyi$^{1,3}$
              and
              G.~Verth}$^{1}$\\[4ex]
  \begin{minipage}[t]{15cm}
        $^1$ Solar Plasma Physics Research Centre,
          School of Mathematics and Statistics,
          University of Sheffield,
          Hounsfield Road, Hicks
          Building, Sheffield, S3 7RH, UK\\
          $^2$ Department of Automatic Control and Systems Engineering, University of Sheffield,
                 Mappin Street, Amy Johnson Building,
                Sheffield, S1 3JD,%
                UK\\
          $^3$ Debrecen Heliophysical Observatory (DHO), Research Centre for Astronomy and Earth Sciences,
Hungarian Academy of Sciences, Debrecen, P.O.Box 30, H-4010, Hungary\\

  {\bf Abstract.}

 Observations suggest that twisted magnetic flux tubes are ubiquitous in the Sun's atmosphere. The main aim of this work is to advance the study of axisymmetric modes of magnetic flux tubes by modeling both twisted internal and external magnetic field, when the magnetic twist is weak. In this work, we solve the derived wave equations numerically assuming twist outside the tube is inversely proportional to the distance from its boundary. We also study the case of constant magnetic twist outside the tube and solve these equations analytically. We show that the solution for a constant twist outside the tube is a good approximation to the case where the magnetic twist is proportional to $1/r$, namely the error is in all cases less than $5.4\%$.The solution is in excellent agreement with solutions to simpler models of twisted magnetic flux tubes, i.e. without external magnetic twist. It is shown that axisymmetric Alfv\'{e}n waves are naturally coupled with magnetic twist as the azimuthal component of the velocity perturbation is nonzero. We compared our theoretical results with observations and comment on what the Doppler signature of these modes is expected to be. Lastly, we argue that the character of axisymmetric waves in twisted magnetic flux tubes can lead to false positives in identifying observations with axisymmetric Alfv\'{e}n waves.

  \vspace*{2ex}
  \end{minipage}
\end{center}
}]


\section{Introduction}
There is ample evidence of twisted magnetic fields in the solar atmosphere and below. For instance, it has been suggested that magnetic flux tubes are twisted whilst rising through the convection zone \citep[see for example][]{murray2008emerging,hood2009emergence,luoni2011twisted}. \cite{brown2003observations,yan2007rapid,kazachenko2009sunspot} have shown that sunspots exhibit a relatively uniform rotation which in turn twists the magnetic field lines emerging from the umbra. Several studies argue that the chromosphere is also permeated by structures that appear to exhibit torsional motion \citep{de2012ubiquitous,sekse2013interplay}. These structures, known as type II spicules, were initially identified by \cite{de2007tale}. \cite{de2012ubiquitous} show that spicules exhibit a dynamical behavior that has three characteristic components, i) flows aligned to the magnetic field, ii) torsional motion and iii) what the authors describe as swaying motion. Also, recent evidence shows that twist and Alfv\'{e}n waves present an important mechanism of energy transport from the photosphere to the corona \citep{wedemeyer2012magnetic}. The increasing body of observational evidence of magnetic twist in the solar atmosphere, in combination with ubiquitous observations of sausage waves \citep{morton2012observations}, reinforce the importance of refining our theoretical understanding of waves in twisted magnetic and especially axisymmetric modes as these could be easily perceived as torsional Alfv\'{e}n waves.

Early studies of twisted magnetic flux tubes focused on stability analyses. For example, \cite{shafranov1956stability} investigated the stability of magnetic flux tubes with azimuthal component of magnetic field proportional to $r$ inside the cylinder and no magnetic twist outside. \cite{kruskal1958hydromagnetic} derived approximate solutions for magnetic flux tubes with no internal twist embedded in an environment with $B_{\varphi} \propto 1/r$. \cite{bennett1999waves} obtained solutions for the sausage mode for stable uniformly twisted magnetic flux tubes with no external twist and \cite{erdelyi2006sausage} extended the analysis for the uncompressible case of constant twist outside the flux tube. The authors also examined the impact of twist on the oscillation periods in comparison to earlier studies \citep[e.g.][]{edwin1983wave} considering magnetic flux tubes with no twist. In a subsequent work \cite{erdelyi2007linear} extended their results in \cite{erdelyi2006sausage} to the compressible case for the sausage mode with no twist outside the tube. \cite{karami2010effect} investigated modes in incompressible flux tubes. The twist was considered to be $\propto r$ for all $r$, which is unphysical for $r \rightarrow \infty$, while the density profile considered was piecewise constant with a linear function connecting the internal and external densities. The authors revealed that the wave frequencies for the kink and fluting modes are directly proportional to the magnetic twist. Also, the bandwidth of the fundamental kink body mode increases proportionally to the magnetic twist.  \cite{terradas2012transverse} investigated twisted flux tubes with magnetic twist localized within a toroidal region of the flux tube and zero everywhere else. \cite{terradas2012transverse} argue that for small twist the main effect of standing oscillations is the change in polarization of the velocity perturbation in the plane perpendicular to the longitudinal dimension ($z$-coordinate).

In this work we study axisymmetric modes, namely eigenmodes corresponding to $\kwvTh=0$, where $\kwvTh$ is the azimuthal wavenumber in cylindrical geometry\footnote{$\kwvTh$ is often denoted as $m$ in a number of other works.}. The azimuthal magnetic field inside the tube is $\propto r$ , while the azimuthal field outside is constant. If there is a current along the tube, according to the Biot-Savart law this current will give rise to a twist proportional to $r$ inside the flux tube and a twist inversely proportional to $r$ outside. For this reason we start our analysis by assuming a magnetic twist outside the tube proportional to $1/r$ and we insert a perturbation parameter that can be used to revert to the case with constant twist. Subsequently, we present an exact solution for the case with constant twist outside the tube and solve numerically for the case with magnetic twist proportional to $1/r$. Then, we compare the numerical solution for $1/r$ with the exact solution for constant twist. Based on the obtained estimated standard error the solution corresponding to weak constant twist appears to be a good approximation to the solution with weak magnetic twist that is proportional to $1/r$. In the case where there is a pre-existing twist inside the cylinder, assuming this twist is uniform, this will give rise to a current which in turn will create the external twist that is again inversely proportional to the distance from the cylinder. The latter case may occur, for example, due to vortical foot-point motions on the photosphere \citep{ruderman1997direct}. Recent observational evidence \citep[e.g.][]{morton2013evidence} put the assumptions of \cite{ruderman1997direct} on a good basis, however, the vortical motions in \cite{morton2013evidence} are not divergence free which means that the same mechanism can be responsible for the axisymmetric modes studied in this work.

The rest of this paper is organized as follows. In \sref{sec:linearized:mhd:equations} we describe the model geometry and MHD equations employed. In \sref{sec:dispersion:equation} we derive the dispersion relation for $\kwvTh=0$, and, in \sref{sec:limiting:cases} we explore limiting cases connecting the results in this work with previous models. Furthermore, in \sref{sec:dispersion:equation:solutions} we study a number of physically relevant cases and elaborate on the results. In, \sref{sec:discussion} we reflect on the applicability and potential limitations of the presented model and in \sref{sec:conclusions} we summarize and conclude this work.

\section{Model Geometry and Basic Equations}\label{sec:linearized:mhd:equations}
\begin{figure}
\includegraphics[width=\columnwidth]{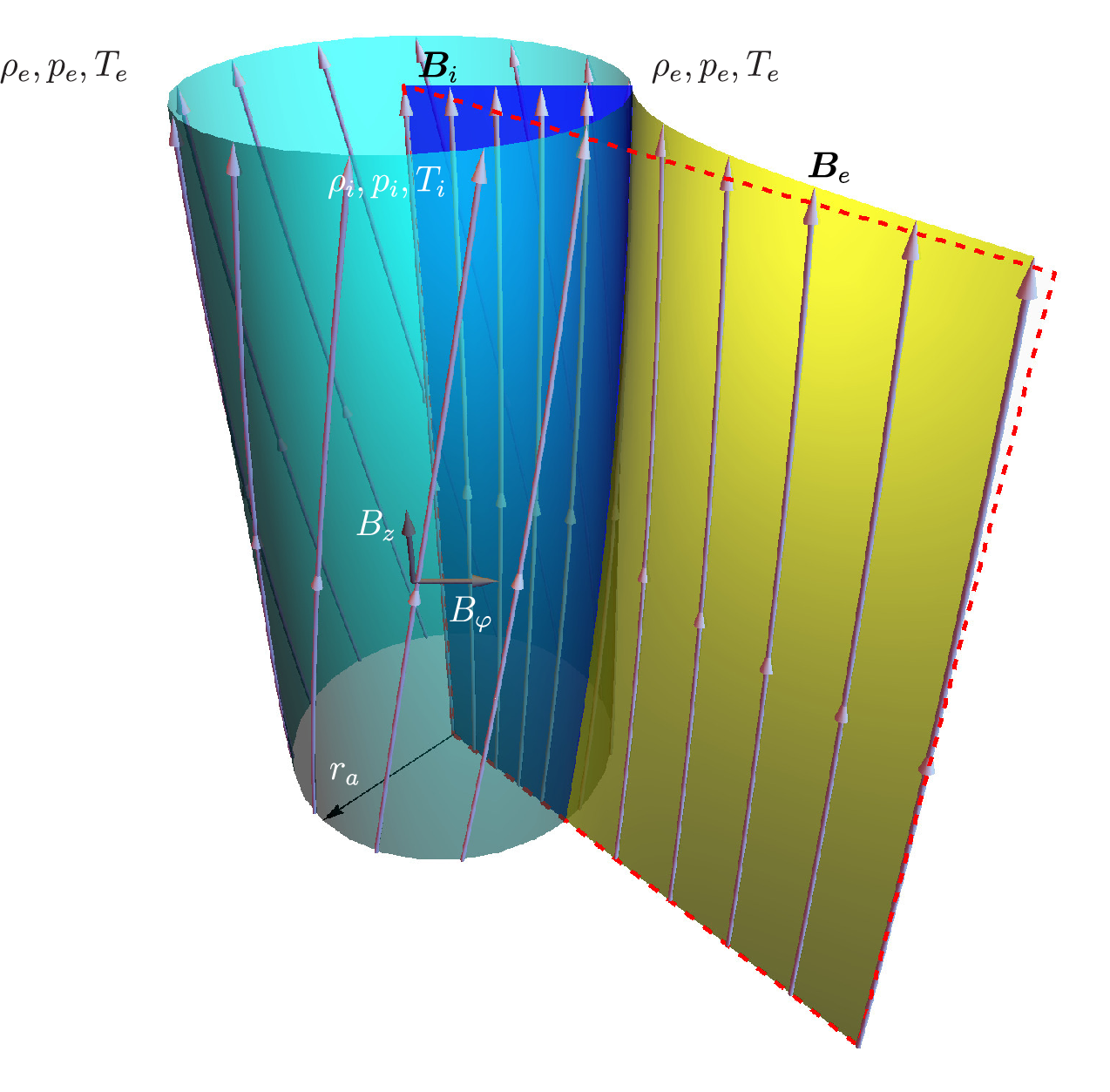}
\caption{Model illustration. Straight magnetic cylinder with variable twist inside ($r<\tuberadius$) and outside ($r>\tuberadius$), where $\tuberadius$ is the tube radius, $\rhoi, \peqi$ and $T_i$ are the density, pressure and temperature at equilibrium inside the tube. The corresponding quantities outside the tube are denoted with a subscript $e$. $\Beqphi$ is continuous across the tube boundary. The dark blue surface inside the magnetic cylinder represents the influence of $\Beqphi \propto r$. The yellow surface outside the cylinder illustrates the $\Beqphi \propto 1/r$ dependence. The dashed red rectangle represents a magnetic surface with only a longitudinal ($z$) magnetic field component.}\label{fig:tube:twist}
\end{figure}

The single-fluid linearized ideal MHD equations in the force formalism are \citep{kadomtsev1966hydromagnetic},
\begin{dgroup}\label{eqn:linearized:mhd}
  \begin{dmath}
\rhoeq \pderiv[2]{\Xip}{t} + \nabla \pp + \frac{1}{\mu_0}\left( \Bp \times (\nabla \times \Beq) + \Beq \times (\nabla \times \Bp) \right) = 0,
  \end{dmath}
\begin{dmath}
\pp + \Xip \cdot \nabla \peq + \gamma \peq \nabla \cdot \Xip = 0,
\end{dmath}
\begin{dmath}
\Bp + \nabla \times (\Beq \times \Xip) = 0,
\end{dmath}
\end{dgroup}
where, $\rhoeq, \peq$ and $\Beq$ are the density, kinetic pressure and magnetic field, respectively, at equilibrium, $\Xip$ is the Lagrangian displacement vector, $\pp$ and $\Bp$ are the pressure and magnetic field perturbation, respectively, $\gamma$ is the ratio of specific heats (taken to be $5/3$ in this work), and $\mu_0$ is the permeability of free space. In this study we employ cylindrical coordinates $(r,\varphi,z)$ and therefore $\Xip = (\Xirp, \Xiphip, \Xizp)$, $\Bp = (\Brp, \Bphip, \Bzp)$. In what follows an index, $i$, indicates quantities inside the flux tube ($r<\tuberadius$) while variables indexed by, $e$, refer to the environment outside the flux tube ($r>\tuberadius$). The model geometry is illustrated in \fref{fig:tube:twist} when $\Beqphie \propto 1/r$. For static equilibrium,
\begin{dgroup}\label{eqn:equilibrium}
\begin{dmath}\label{eqn:equil:cond1}
  \nabla \cdot \Beq = 0,
\end{dmath}
\begin{dmath}\label{eqn:equil:cond2}
  \Jeq = \frac{1}{\mu_0} \nabla \times \Beq,
\end{dmath}
\begin{dmath}\label{eqn:equil:cond3}
  \nabla \peq = \Jeq \times \Beq.
\end{dmath}
\end{dgroup}
We assume that, $\rhoeq,\peq$ and $\Beq$ have only an $r$-dependence. We consider a magnetic field of the following form,
\begin{dmath}
  \Beq = (0,\Beqphi(r),\Beqz(r)).
\end{dmath}
Notice that in cylindrical coordinates, \eref{eqn:equil:cond1} is identically satisfied. Then, \eref{eqn:equil:cond2} expands to
\begin{dmath}\label{eqn:equil:cond2:exp1}
  \mu_0 \Jeq = \left(0, - \D{\Beqz}{r}, \frac{1}{r}\D{(r \Beqphi)}{r} \right),
\end{dmath}
and based on \eref{eqn:equil:cond2:exp1}, \eref{eqn:equil:cond3} becomes
\begin{dmath}
  \nabla \peq = \left(-\frac{\Beqz}{\mu_0} \pderiv[1]{\Beqz}{r} - \frac{\Beqphi}{\mu_0 r} \D{(r \Beqphi)}{r},0, 0\right).
\end{dmath}
Therefore, the pressure in the $\varphi$ and $z$ directions is constant and the magnetic field and the plasma pressure must satisfy the following pressure balance equation in the $r$ direction,
\begin{dmath}\label{eqn:pressure:r}
 \D{}{r}\!\left(\peq + \frac{\Beqphi^2 + \Beqz^2}{2 \mu_0} \right) + \frac{\Beqphi^2}{\mu_0 r} = 0.
\end{dmath}
For a magnetic flux tube of radius, $\tuberadius$, according to the Biot-Savart law (for $\mee = 1$ in \eref{eqn:bfield:in:out}) a reasonable assumption for the form of the magnetic field is,
\begin{dmath}\label{eqn:bfield:in:out}
  \Beq = \twopartdef{\left(0, \TwistConstI r, \Beqzi\right)}{r \leq \tuberadius,}{\left(0, r_{a}^{1+\mee}\TwistConstE /r^{\mee},\Beqze\right)}{r > \tuberadius,}
\end{dmath}
where, $\Beqzi, \Beqze$, $\TwistConstI$ and $\TwistConstE$ are constants and $\mee$ is a perturbation parameter. The perturbation parameter has been inserted in \eref{eqn:bfield:in:out} in such a way so as to preserve dimensional consistency. The constant $\TwistConstI$ can be determined by application of the Biot-Savart law and is therefore taken to be,
\begin{dmath}\label{eqn:biot:savart}
  \TwistConstI = \frac{\mu_0 I}{2 \pi \tuberadius^{1+\mee}},
\end{dmath}
where $I$ is the current. By substituting \eref{eqn:bfield:in:out} into \eref{eqn:pressure:r} we obtain:
\begin{dmath}\label{eqn:pressure:pr}
\displaystyle p(r) = \twopartdef{\displaystyle  \frac{\TwistConstI^2}{\mu_0}(\tuberadius^2 - r^2) + p_a } {r \leq \tuberadius,} {\displaystyle  \frac{\tuberadius^{2(1+\mee)} \TwistConstE^2(1- 2 \mee)}{2 \mu_0 \mee} \left( \frac{1}{r^{2 \mee}} - \frac{1}{\tuberadius^{2\mee}}\right) + p_a } {r>\tuberadius,}
\end{dmath}
where, $p_a$, is the pressure at the boundary of the magnetic flux tube. The constant, $\TwistConstE$, is equal to $\TwistConstI$, however we choose to maintain the notational distinction so that we can separate the internal and external environments to the flux tube which helps us validate our results with previous work \citep[e.g.][]{erdelyi2007linear}.

\subsection{Governing Equations}\label{sec:governing:equations}
The solution of the system of equations shown in \eref{eqn:linearized:mhd}, in cylindrical coordinates can be found by Fourier decomposition of the perturbed components, namely the perturbed quantities are taken to be,
\begin{dmath}\label{eqn:fourier:xi:pt}
\Xip, \pTp \propto e^{i\left(\kwvTh \varphi + \kwvZ z - \omega t\right)},
\end{dmath}
where, $\omega$ is the angular frequency, $\kwvTh$ is the azimuthal wavenumber for which only integer values are allowed and, $\kwvZ$, is the longitudinal wavenumber in the $z$ direction. The Eulerian total pressure perturbation is, $\pTp = \pp + \Beq \Bp/\mu_0$, which is obtained by linearization of the total pressure: $p_T = (\peq + \pp) + (\Beq + \Bp)^2 / (2\mu_0)$ and $\peq$ is the equilibrium kinetic pressure. Note that for the sausage mode, considered in this work, the azimuthal wavenumber is $\kwvTh = 0$. Combining \eref{eqn:linearized:mhd} with \eref{eqn:fourier:xi:pt} we obtain the equation initially derived by \cite{hain1958normal} and later by \cite{goedbloed1971stabilization,sakurai1991resonantI} to name but a few. This equation can be reformulated as two coupled first order differential equations,
\begin{dgroup}\label{eqn:sakurai}
\begin{dmath}\label{eqn:sakurai:xidel}
D \D{(r \Xirp)}{r} = C_1 (r \Xirp) - r C_2 \pTp,
\end{dmath}
\begin{dmath}\label{eqn:sakurai:ptdel}
D \D{\pTp}{r} = \frac{1}{r} C_3 (r \Xirp) - C_1 \pTp.
\end{dmath}
\end{dgroup}
and the multiplicative factors are defined as:
\begin{dgroup}\label{eqn:parameters}
\begin{dmath}
D = \rho (\omega^2 - \omega_A^2) C_4,
\end{dmath}
\begin{dmath}
C_1 = \frac{2 \Beqphi}{\mu_0 r} \left( \omega^4 \Beqphi - \frac{\kwvTh}{r} f_B C_4 \right),
\end{dmath}
\begin{dmath}
C_2 = \omega^4 - \left(\kwvZ^2 + \frac{\kwvTh^2}{r^2}\right)C_4,
\end{dmath}
\begin{dmath}
C_3 = \rho D \left[ \omega^2 - \omega_A^2 + \frac{2 \Beqphi}{\mu_0 \rho} \D{}{r}\left( \frac{\Beqphi}{r} \right) \right] \\+ 4 \omega^4\frac{ \Beqphi^4}{\mu_0^2 r^2} - \rho C_4 \frac{ 4 \Beqphi^2 \omega_A^2 }{\mu_0 r^2},
\end{dmath}
\begin{dmath}
C_4 = (\vS^2 + \vA^2)(\omega^2 - \omega_c^2),
\end{dmath}
\end{dgroup}
where,
\begin{align*}
\omegaC^2 = \frac{\vS^2}{\vA^2 + \vS^2} \omegaA^2, && \omegaA^2 = \frac{f_B^2}{\mu_0 \rhoeq}, && f_B = \frac{\kwvTh}{r} \Beqphi + \kwvZ \Beqz.
\end{align*}
Here, $\vS = \sqrt{\gamma \peq / \rhoeq}$ is the sound speed, $\vA = |\Beq|/\sqrt{\mu_0 \rhoeq}$ is the Alfv\'{e}n speed, $\omegaC$ is the cusp angular frequency and $\omegaA$ is the Alfv\'{e}n angular frequency. The coupled first order ODEs in \eref{eqn:sakurai} can be combined in a single second order ODE for, $\pTp$ or $\Xirp$. In this work we choose to use the latter approach, namely:
\begin{dmath}\label{eqn:sakurai:ode}
\D{}{r}\left[ \frac{D}{r C_2} \D{}{r}(r \Xirp)\right] + \left[ \frac{1}{D}\left(C_3 - \frac{C_1^2}{C_2} \right) - r \D{}{r}\left( \frac{C_1}{r C_2} \right) \right] \Xirp = 0.
\end{dmath}
Using flux coordinates and assuming $\kwvTh = 0$, it can be shown that \citep{sakurai1991resonantI},
\begin{dgroup}\label{eqn:xi:flux}
\begin{dmath}\label{eqn:xi:flux:perp}
\rhoeq (\omega^2 - \omegaA^2) \Xiperp = - i \frac{\kwvZ \Beqphi}{| \Beq |} \left( \pTp + 2 \frac{\Beqz^2}{\mu_0 r} \Xirp \right),
\end{dmath}
\begin{dmath}\label{eqn:xi:flux:para}
\rhoeq (\omega^2 - \omega_c^2) \Xipara = i \frac{\kwvZ \Beqz}{| \Beq |} \frac{\vS^2}{\vS^2 + \vA^2}  \left( \pTp - 2 \frac{\Beqphi^2}{\mu_0 r} \Xirp \right).
\end{dmath}
\end{dgroup}
Here $\Xipara$ and $\Xiperp$ are the Lagrangian displacement components parallel and perpendicular to the magnetic field lines respectively. The dominant component of the Lagrangian displacement vector $(\Xirp, \Xiperp, \Xipara)$ determines the character of the mode. For the Alfv\'{e}n mode the $\Xiperp$ component is dominant, while for the slow and fast magnetoacoustic modes $\Xipara$ and $\Xirp$ is dominant respectively \citep{goossens2011resonant}.  \eref{eqn:xi:flux} suggests that in the presence of magnetic twist the slow and fast magnetoacoustic modes are coupled to the Alfv\'{e}n mode even when $\kwvTh = 0$, namely the slow and fast modes do not exist without the Alfv\'{e}n mode and vice versa. This is because for the Alfv\'{e}n mode to be decoupled from the slow and fast magnetoacoustic modes it is required that for $\Xiperp \neq 0$, $\Xirp = 0$ and $\Xipara = 0$. However, it follows trivially from \eref{eqn:sakurai} that, if $\Xirp = 0$ then also $\pTp = 0$ and therefore from \eref{eqn:xi:flux} we have that $\Xiperp = 0$. From this, it follows that the Alfv\'{e}n mode cannot exist without the components corresponding to the slow and fast magnetoacoustic modes, hence the Alfv\'{e}n mode is coupled with the slow and fast magnetoacoustic modes. Furthermore, from \eref{eqn:xi:flux} we can also see that for a solution, i.e. $(\omega, \kwvZ)$ pair, as $\omega$ approaches $\omegaA$ the $\Xiperp$ component is amplified that leads to the azimuthal component of the displacement to be accentuated.

\section{Dispersion Equation}\label{sec:dispersion:equation}
In this section we follow a standard procedure in deriving a dispersion equation, namely we solve \eref{eqn:sakurai:ode} inside and outside the flux tube and match the two solutions using the boundary conditions. The boundary conditions that must be satisfied are:
\begin{dgroup}
\begin{dmath}\label{eqn:displacement:continuity}
\left.\Xirip \right\vert_{ r = \tuberadius} = \left.\Xirep \right\vert_{ r = \tuberadius },
\end{dmath}
\begin{dmath}\label{eqn:total:pressure:continuity}
\pTip - \left.\frac{\Bphii^2}{\mu_0 r} \Xirip \right\vert_{ r = \tuberadius } = \pTep - \left. \frac{\Bphie^2}{\mu_0 r} \Xirep\right\vert_{ r = \tuberadius },
\end{dmath}
\end{dgroup}
where, \eref{eqn:displacement:continuity} and \eref{eqn:total:pressure:continuity} are continuity conditions for the Lagrangian displacement and total pressure across the tube boundary respectively.

\subsection{Solution Inside the Flux Tube}
The parameters in \eref{eqn:parameters} for the case inside the flux tube for the sausage mode become,
\begin{dgroup}\label{eqn:parameters:inside}
\begin{dmath}
D_i = \rhoi (\omega^2 - \omegaAi^2),
\end{dmath}
\begin{dmath}
C_1 = \frac{2 \TwistConstI^2 r}{\mu_0} \Ni^2,
\end{dmath}
\begin{dmath}
C_2 = n_i^2 - \kwvZ^2,
\end{dmath}
\begin{dmath}
C_3 = \rhoi \left[ (\omega^2 - \omegaAi^2)^2 + \frac{ 4 \TwistConstI^2}{\mu_0 \rhoi} \left( \frac{ \TwistConstI^2 r^2}{\mu_0 \rhoi} \Ni^2 - \omegaAi^2 \right) \right],
\end{dmath}
\begin{dmath}
\Ni^2 = \frac{ \omega^4 }{ (\vSi^2 + \vAi^2)(\omega^2 - \omegaCi^2) },
\end{dmath}
\end{dgroup}
where,
\begin{align*}
\omegaCi^2 = \frac{\vSi^2}{\vAi^2 + \vSi^2} \omegaAi^2, && \omegaAi^2 = \kwvZ^2 \frac{\Beqzi^2}{\mu_0 \rhoi}.
\end{align*}
The $\Beqphi$ component is assumed small to avoid the kink instability, see for example \citep{gerrard2002triggering,torok2003ideal}. This implies, $\Beqphi \ll \Beqz$ and since $\Beqphi$ is a function of $r$ inside (and outside) the tube we require $\sup(\Beqphi) \ll \Beqz \Rightarrow \TwistConst \tuberadius \ll \Beqz$. This condition is satisfied in the solar atmosphere, so we can use the approximation: $\vAi^2 = (\Beqphii^2 + \Beqzi^2)/(\mu_0 \rhoi) \sim \Beqzi^2/(\mu_0 \rhoi)$. Notice that according to \eref{eqn:pressure:pr} the pressure depends on $r$, however, in this work we assume that the sound speed is constant. This is because the term that depends on $r$ in \eref{eqn:pressure:pr} is assumed to be small when compared with $p_a$ in solar atmospheric conditions. To see this, consider that $\sup( \TwistConst \tuberadius ) = 0.2 \Beqz$ and\footnote{In the following expressions the left number corresponds to typical values on the photosphere while the right number corresponds to typical values of the quantity in the corona.} $\Beqz \sim (10^{-1} - 10^{-4})T$, $T \sim (10^{4} - 10^{6})K$ and the number density $n \sim (10^{23} - 10^{16})m^{-3}$. This means that\footnote{Here we use $p = n k_B T$.} $p_a \sim (10^{4} - 10^{-1}) N \cdot m^{-2}$ and the term that depends on the radius is of the order $(\TwistConst \tuberadius)^{2} / \mu_0 \sim (10^{2} - 10^{-4}) N \cdot m^{-2}$ and therefore the constant term $p_a$ is $(10^2 - 10^3)$ times larger when compared with the term that has an $r$ dependence. Hence, to a good approximation, the pressure can be assumed to be constant. Note, that the density is discontinuous across the tube boundary and therefore we avoid the Alfv\'{e}n and slow continua that lead to resonant absorption.

Substitution of the parameters in \eref{eqn:parameters:inside} into \eref{eqn:sakurai:ode}, leads to the following second order differential equation \citep[see for example][]{erdelyi2007linear},
\begin{dmath}\label{eqn:xirode:inside}
R^2 \D{^2 \Xirp}{R^2} + R \D{\Xirp}{R} - \left(1 + \frac{\kwvRin^2}{\kwvZ^2}R^2 + E R^4\right) \Xirp = 0,
\end{dmath}
where,
\begin{dgroup*}
\begin{dmath*}
R = k_{\alpha} r,
\end{dmath*}
\begin{dmath*}
\kwvRin^2 = \frac{(\kwvZ^2 \vSi^2 - \omega^2)(\kwvZ^2 \vAi^2 - \omega^2)}{(\vAi^2 + \vSi^2)(\kwvZ^2 \vTi^2 - \omega^2)},
\end{dmath*}
\begin{dmath*}
E = \frac{4 \TwistConstI^4 \Ni^2}{\mu_0^2 D_i^2 \kwvZ^2 (1-\alpha^2)^2},
\end{dmath*}
\begin{dmath*}
\alpha^2 = \frac{4 \TwistConstI^2 \omegaAi^2}{\mu_0 \rhoi (\omega^2 - \omegaAi^2)^2},
\end{dmath*}
\begin{dmath*}
\vTi^2 = \frac{\vAi^2 \vSi^2}{\vAi^2 + \vSi^2}.
\end{dmath*}
\end{dgroup*}
Here $k_{\alpha} = \kwvZ(1-\alpha^2)^{1/2}$ is the effective longitudinal wavenumber and $\vTi$ is the internal tube speed. \eref{eqn:xirode:inside} was derived and solved before by \cite{erdelyi2007linear}. The solution is expressed in terms of Kummer functions \citep{abramowitz2012handbook} as follows,
\begin{dmath}
\Xirp(\kummerVar) = A_{i 1} \frac{\kummerVar^{1/2}}{E^{1/4}} e^{-\kummerVar/2} \KummerM{\kummerA}{\kummerB}{\kummerVar} + A_{i 2} \frac{\kummerVar^{1/2}}{E^{1/4}} e^{-\kummerVar/2} \KummerU{\kummerA}{\kummerB}{\kummerVar},
\end{dmath}
and the parameters, $\kummerA,\kummerB$ and the variable $\kummerVar$ are defined as
\begin{dgroup*}
\begin{dmath*}
\kummerA = 1 + \frac{\kwvRin^2}{4 \kwvZ^2 E^{1/2}},
\end{dmath*}
\begin{dmath*}
\kummerB = 2,
\end{dmath*}
\begin{dmath*}
\kummerVar = R^2 E^{1/2} = k_{\alpha}^2 E^{1/2} r^2,
\end{dmath*}
\end{dgroup*}
$A_{i 1}$ and $A_{i 2}$ are constants. Furthermore, the total pressure perturbation, $\pTp$ is:
\begin{dmath*}
\pTp(\kummerVar) = A_{i 1} \frac{k_a D_i}{\Ni^2 - \kwvZ^2} e^{-\kummerVar/2}\left[ \frac{\Ni+\kwvZ}{\kwvZ} \kummerVar \KummerM{\kummerA}{\kummerB}{\kummerVar} - 2 \KummerM{\kummerA}{\kummerB-1}{\kummerVar}\right] \\ + A_{i 2}\frac{k_a D_i}{\Ni^2-\kwvZ^2} e^{-\kummerVar/2} \left[\frac{\Ni+\kwvZ}{\kwvZ} \kummerVar \KummerU{\kummerA}{\kummerB}{\kummerVar} - 2(1-\kummerA)\KummerU{\kummerA}{\kummerB-1}{\kummerVar}\right].
\end{dmath*}
Now, considering that solutions at the axis of the flux tube, namely at $r=0$, must be finite, we take $A_{i 2} = 0$ and so
\begin{dgroup}
\begin{dmath}\label{eqn:xir:inside:solution}
\Xirip(\kummerVar) = A_{i 1} \frac{\kummerVar^{1/2}}{E^{1/4}} e^{-\kummerVar/2} \KummerM{\kummerA}{\kummerB}{\kummerVar},
\end{dmath}
\begin{dmath}\label{eqn:pti:inside:solution}
\pTip(\kummerVar) = A_{i 1} \frac{k_a D_i}{\Ni^2 - \kwvZ^2} e^{-\kummerVar/2}\left[ \frac{\Ni+\kwvZ}{\kwvZ} \kummerVar \KummerM{\kummerA}{\kummerB}{\kummerVar} - 2 \KummerM{\kummerA}{\kummerB-1}{\kummerVar} \right].
\end{dmath}
\end{dgroup}
Note that the corresponding equation to \eref{eqn:pti:inside:solution} had a typographical error in \cite{erdelyi2007linear} (see Equation (13) in that work).

\subsection{Solution Outside the Flux Tube}
The multiplicative factors in \eref{eqn:parameters} outside of the flux tube for the sausage mode, $\kwvTh=0$, become
\begin{dgroup}\label{eqn:parameters:outside}
\begin{dmath}
D_e = \rhoe (\omega^2 - \omegaAe^2),
\end{dmath}
\begin{dmath}
C_1 = \frac{2 \tuberadius^{2(1+\mee)} \TwistConstE^2}{\mu_0 r^{1+2 \mee}} \Ne^2,
\end{dmath}
\begin{dmath}
C_2 = \Ne^2 - \kwvZ^2,
\end{dmath}
\begin{dmath}
C_3 = \rhoe^2 (\omega^2 - \omegaAe^2)^2  \\
+\frac{4 \tuberadius^{2(1+\mee)} \TwistConstE^2}{\mu_0 r^{2(1+\mee)}} \left[ \frac{\tuberadius^{2(1+\mee)} \TwistConstE^2}{\mu_0 r^{2 \mee}} n_e^2 - \rhoe \omegaAe^2 - \rhoe\frac{(1+\mee)}{2}(\omega^2 - \omegaAe^2)\right],
\end{dmath}
\begin{dmath}
\Ne^2 = \frac{ \omega^4 }{ (\vSe^2 + \vAe^2)(\omega^2 - \omegaCe^2) },
\end{dmath}
\end{dgroup}
where,
\begin{align*}
\omegaCe^2 = \frac{\vSe^2}{\vAe^2 + \vSe^2} \omegaAe^2, && \omegaAe^2 = \kwvZ^2 \frac{\Beqze^2}{\mu_0 \rhoe}.
\end{align*}
\eref{eqn:sakurai:ode} with \eref{eqn:parameters:outside} for $\mee = 1$ corresponds to $\Beqphi \sim 1/r$, however, the resulting ODE is difficult to solve. By setting $\mee = 0$ we obtain the case for constant twist outside the tube, which is also a zeroth-order approximation to the problem with $\mee = 1$ \citep{bender1999advanced}. Note, that it is unconventional to use only the zeroth-order term in perturbative methods, and therefore, to establish the validity of the approximation we estimate the error by solving for $\mee = 1$ numerically. The estimated error is quoted in the caption of the dispersion diagrams in this work and the process which we followed to obtain this is described in \axref{apx:rmse}. Substituting the parameters given in \eref{eqn:parameters:outside} into \eref{eqn:sakurai:ode} we have
\begin{dmath}\label{eqn:xirode:outside}
r^2 \D{^2 \Xirp}{r^2} + r \D{\Xirp}{r} - \left( \kwvRout^2 r^2 + \nu^2(\mee ; r) \right)\Xirp = 0,
\end{dmath}
where, $\kwvRout^2$ and $\nu^2$:
\begin{dgroup*}
\begin{dmath}
\kwvRout^2 = -(\Ne^2 - \kwvZ^2),
\end{dmath}
\begin{dmath}\label{eqn:nu:square:twist}
\nu^2(\mee; r) = 1 + 2\frac{\tuberadius^{2(1+\mee)} \TwistConstE^2}{\mu_0^2 D_e^2 r^{2\mee}}\left\{ 2\frac{\tuberadius^{2(1+\mee)} \TwistConstE^2 \Ne^2 \kwvZ^2}{r^{2 \mee}} + \mu_0 \rhoe \left[\omegaAe^2(n_e^2(3+\mee)\\-\kwvZ^2(1-\mee)) - (\Ne^2 + \kwvZ^2)(1+\mee)\omega^2 \right]\right\}.
\end{dmath}
\end{dgroup*}
Notice that $\nu^2(0 ; r)$ is independent of $r$. Therefore, for $\mee \rightarrow 0$, \eref{eqn:xirode:outside} is transformed to either the Bessel equation for $\kwvRout^2 <0$ or the modified Bessel equation for $\kwvRout^2 > 0$. It should be noted that $\Ne^2 = \kwvZ^2$, namely $\kwvRout^2 = 0$, is prohibited since during the derivation of \eref{eqn:xirode:outside} it was assumed that $\Ne^2 \neq \kwvZ^2$ to simplify the resulting equation. Therefore, the solution to \eref{eqn:xirode:outside} for $\mee \rightarrow 0$, and, assuming no energy propagation away from or towards the cylinder ($\kwvRout^2 > 0$), is
\begin{dmath}
\Xirp(r) = A_{e 1}K_{\nu}\left(\kwvRout r\right)+A_{e 2}I_{\nu}\left(\kwvRout r\right).
\end{dmath}
On physical grounds we require the solution to be evanescent, i.e. $\Xirp(r) \rightarrow 0$ as $r \rightarrow 0$, and therefore we must have $A_{e 2}=0$, namely
\begin{dmath}\label{eqn:xir:outside:solution}
\Xirep(r) = A_{e 1}K_{\nu}\left(\kwvRout r\right),
\end{dmath}
and, from \eref{eqn:sakurai:xidel}, the total pressure perturbation $\pTep$ is
\begin{dmath}\label{eqn:pte:outside:solution}
\pTep =
A_{e 1} \left(\frac{ \mu_0 (1-\nu) D_e - 2 \tuberadius^{2} \TwistConstE^2 n_e^2 }{\mu_0 r (\kwvZ^2 - \Ne^2)} K_{\nu}(\kwvRout r) - \frac{ D_e }{\kwvRout} K_{\nu-1}(\kwvRout r) \right).
\end{dmath}
Note that, for the case $\kwvRout^2>0$ and $\TwistConstE \rightarrow 0$, namely zero twist outside the cylinder, $\nu^2 \rightarrow 1$, thus retrieving the solution for $\Xirp$ by \cite{edwin1983wave}. The limiting cases for \eref{eqn:xir:inside:solution} and \eref{eqn:pti:inside:solution} have been verified to converge to the solutions with no twist inside the magnetic cylinder in \cite{erdelyi2007linear} in \sref{sec:limiting:cases}.

\subsection{Dispersion Relation and Limiting Cases}\label{sec:limiting:cases}
Application of the boundary conditions (see \eref{eqn:displacement:continuity} and \eref{eqn:total:pressure:continuity}) in combination to the solutions for $\Xirp$ and $\pTp$ inside the magnetic flux tube, \eref{eqn:xir:inside:solution} and \eref{eqn:pti:inside:solution} as well as the solutions in the environment of the flux tube, \eref{eqn:xir:outside:solution} and \eref{eqn:pte:outside:solution} respectively, leads to the following general dispersion equation, for the compressible case in presence of internal and external magnetic twist,
\begin{dmath}\label{eqn:dispersion:equation}
\frac{\tuberadius D_e}{\kwvRout} \frac{K_{\nu-1}(\kwvRout \tuberadius)}{K_{\nu}(\kwvRout \tuberadius)} = \frac{\tuberadius^2}{\mu_0} \left[ \frac{\TwistConstI^2}{\kwvRin^2}(\Ni + \kwvZ)^2 - \frac{\TwistConstE^2}{\kwvRout^2}(\Ne^2 + \kwvZ^2) \right] + \frac{(1-\nu)D_e}{\kwvRout^2} - 2 \frac{D_i}{\kwvRin^2}\frac{M(\kummerA,\kummerB-1;\kummerVar)}{M(\kummerA,\kummerB;\kummerVar)}.
\end{dmath}
In order to validate \eref{eqn:dispersion:equation} we consider a number of limiting cases. First, the case where there is no external magnetic twist. In this case $\TwistConstE = 0$ and from \eref{eqn:nu:square:twist} it follows trivially that, $\nu^2(\mee; r) = 1$. Therefore, \eref{eqn:dispersion:equation} for no external twist becomes:
\begin{dmath}\label{eqn:dispersion:equation:noexternaltwist}
\frac{\tuberadius D_e}{\kwvRout} \frac{K_{0}(\kwvRout \tuberadius)}{K_{1}(\kwvRout \tuberadius)} = \frac{\TwistConstI^2 \tuberadius^2}{\mu_0 \kwvRin^2}(\Ni + \kwvZ)^2  - 2 \frac{D_i}{\kwvRin^2}\frac{M(\kummerA,\kummerB-1;\kummerVar)}{M(\kummerA,\kummerB;\kummerVar)}.
\end{dmath}
This equation is in agreement with the dispersion equation obtained by \cite{erdelyi2007linear} and all the limiting cases therein also apply to \eref{eqn:dispersion:equation}. However, there is one limiting case missing from the analysis in \cite{erdelyi2007linear}, namely for no twist inside and outside the tube with $\kwvRin^2<0$, which in combination to $\kwvRout^2 > 0$ corresponds to body wave modes. We complete this analysis here. Starting with 13.3.1 and 13.3.2 in \cite{abramowitz2012handbook}, in the limit as $\TwistConstI \rightarrow 0$ and $\kwvRin^2 > 0$ we have
\begin{dgroup}\label{eqn:kummer:limits:m:pos}
\begin{dmath}
\underset{\TwistConstI \rightarrow 0}{\lim} \left(M(\kummerA,\kummerB-1;\kummerVar)\right) =  \BesselI{0}{\kwvRin r},
\end{dmath}
\begin{dmath}
\underset{\TwistConstI \rightarrow 0}{\lim} \left(M(\kummerA,\kummerB;\kummerVar)\right) = \frac{2}{\kwvRin r} \BesselI{1}{\kwvRin r},
\end{dmath}
\end{dgroup}
while for $\kwvRin^2 < 0$:
\begin{dgroup}\label{eqn:kummer:limits:m:neg}
\begin{dmath}
\underset{\TwistConstI \rightarrow 0}{\lim} \left(M(\kummerA,\kummerB-1;\kummerVar)\right) = \BesselJ{0}{\kwvRin r},
\end{dmath}
\begin{dmath}
\underset{\TwistConstI \rightarrow 0}{\lim} \left(M(\kummerA,\kummerB;\kummerVar)\right) = \frac{2}{\kwvRin r} \BesselJ{1}{\kwvRin r}.
\end{dmath}
\end{dgroup}
Therefore, \eref{eqn:dispersion:equation:noexternaltwist} in conjunction with the facts that $J_0^{\prime}(\kummerVar) = -J_1(\kummerVar)$, $I_0^{\prime}(\kummerVar) = I_1(\kummerVar)$ and $K_0^{\prime}(\kummerVar) = -K_1(\kummerVar)$ \citep[9.1.28 and 9.6.27 in][]{abramowitz2012handbook} is, in the limit as $\TwistConstI \rightarrow 0$ equal to
\begin{align}\label{eqn:edwin:roberts:dispersion}
\kwvRin D_e \frac{K_0(\kwvRout \tuberadius)}{K_0^{\prime}(\kwvRout \tuberadius)} = \kwvRout D_i \frac{I_0(\kwvRin \tuberadius)}{I_0^{\prime}(\kwvRin \tuberadius)}, & \text{ for } \kwvRin^2 > 0, \\
\abs{\kwvRin} D_e \frac{K_0(\kwvRout \tuberadius)}{K_0^{\prime}(\kwvRout \tuberadius)} = \kwvRout D_i \frac{J_0(\abs{\kwvRin} \tuberadius)}{J_0^{\prime}(\abs{\kwvRin} \tuberadius)}, & \text{ for } \kwvRin^2 < 0,
\end{align}
which are in agreement with \cite{edwin1983wave} and describe wave mode for the case with no magnetic twist.

\section{Dispersion Equation Solutions}\label{sec:dispersion:equation:solutions}
\begin{table}[t]
\resizebox{\columnwidth}{!}{
 \begin{tabular}{clccc}
 \hline
Characteristic Speeds Ordering & Type & $\beta_i$ & $\beta_e$  \\
 \hline
$\vSE > \vSI > \vAE > \vAI$ & Warm dense tube     & $>1$     & $>1$     \\
$\vSE > \vAI > \vAE > \vSI$ & Cool evacuated tube & $\ll 1$  & $\gg 1$  \\
$\vSE > \vSI > \vAI > \vAE$ & Weak cool tube      & $> 1$    & $\gg 1$  \\
$\vAE > \vSI > \vAI > \vSE$ & Intense warm tube   & $\ll 1$  & $\ll 1$  \\
$\vAE > \vAI > \vSE > \vSI$ & Intense cool tube   & $\ll 1$  & $\ll 1$  \\
 \hline
 \end{tabular}
 }
  \caption{Equilibrium cases considered in this work. The normalized characteristic speeds are defined in \axref{apx:non:dimensional:dispersion}.\label{tbl:cases}}
\end{table}

To explore the behavior of the sausage mode \eref{eqn:dispersion:equation} was normalized and solved numerically for different solar atmospheric conditions, see \tref{tbl:cases}. Normalized quantities are denoted with capitalized indices (see \axref{apx:non:dimensional:dispersion}). The solutions of the dispersion relation depend only on the relative ordering of the magnitudes of the characteristic velocities ($\vAE,\vAI,\vSE,\vSI$). The sign of $\kwvRIN^2$ and $\kwvROUT^2$ depends on this ordering and this in turn defines the three band types in the dispersion plot, i) bands that contain surface modes (for $\kwvRIN^2>0$ and $\kwvROUT^2>0$), ii) bands that contain body modes (for $\kwvRIN^2<0$ and $\kwvROUT^2>0$), and, iii) forbidden bands corresponding to $\kwvROUT^2 < 0$. We make additional comments on the selection of the characteristic speeds in \axref{apx:speeds:ordering}. The non-dimensional dispersion equation is given in \axref{apx:non:dimensional:dispersion} while the solutions for the perturbed quantities in terms of $\Xirp$ and $\pTp$ are given in \axref{apx:perturbed:quantities}.

\subsection{High plasma-$\beta$ regime}\label{sec:high:plasma:beta}
\begin{figure}

\includegraphics[width=\columnwidth]{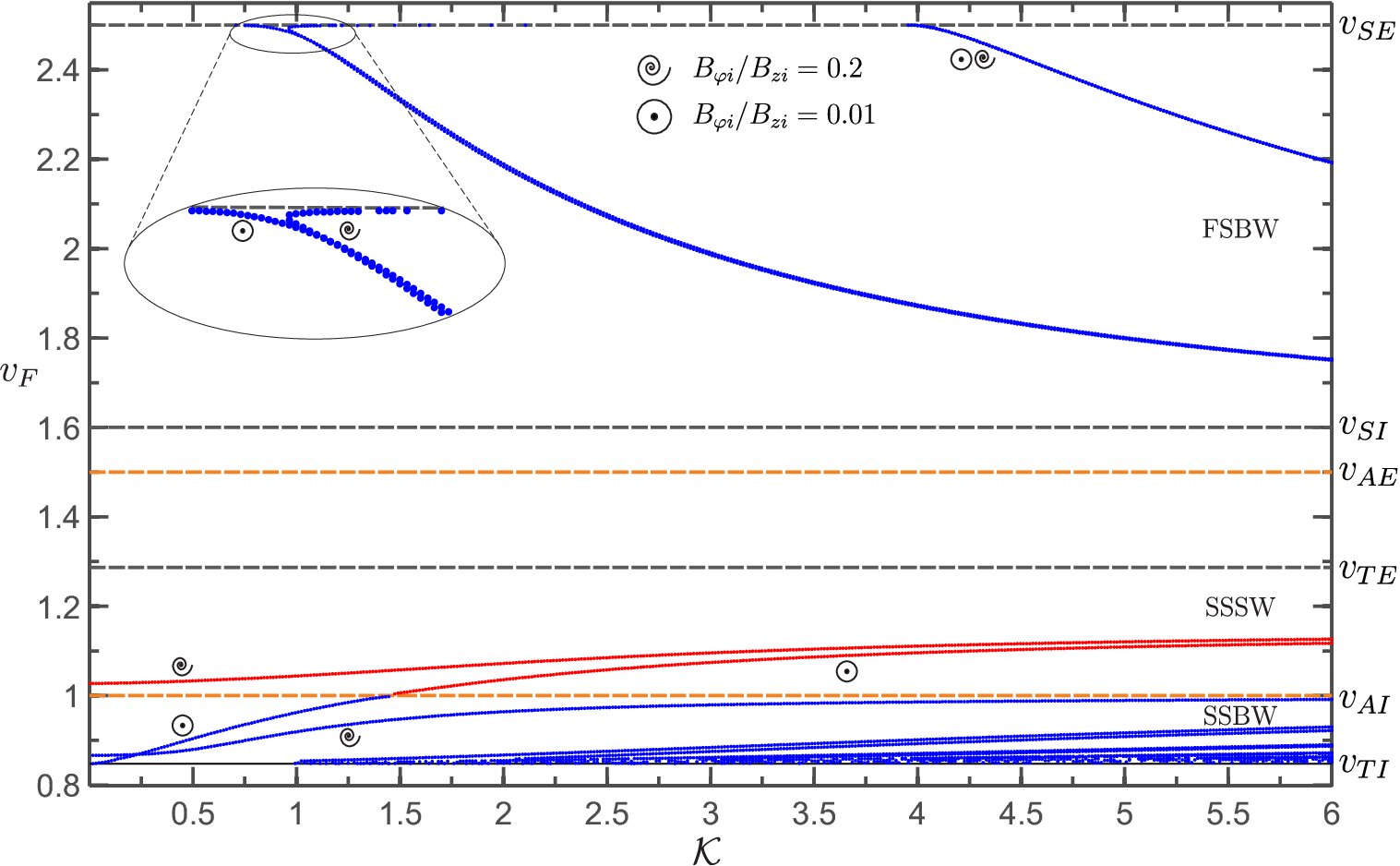}
\caption{Solutions of the dispersion \eref{eqn:dispersion:equation} for a warm dense tube embedded in a dense environment ($\beta_i >1, \beta_e>1$) with speed ordering $\vSE > \vSI > \vAE > \vAI$. The color coding is as follows: blue indicates body waves, red corresponds to surface waves, orange corresponds to either the internal or external Alfv\'{e}n speeds, note this convention is used consistently in this work. The circle with a point corresponds to the case $\Beqphii / \Beqzi = 0.01$ while the spiral corresponds to $\Beqphii / \Beqzi = 0.2$. The mean root mean squared error (RMSE) for this speed ordering is $0.0328$. \label{fig:case:1:dispersion}}
\end{figure}

Based on the results by \cite{vernazza1981structure} and the model for the plasma-$\beta$ in the solar atmosphere introduced by \cite{gary2001plasma} we anticipate that the results in this section are pertinent to conditions typically observed in the upper photosphere, lower chromosphere and mid-chromosphere. The solutions of the dispersion relation in \eref{eqn:dispersion:equation}, in terms of the non-dimensional phase speed, $\vF = \vph / \vAi = \omega / \kwvZ \vAi$, and the non-dimensional longitudinal wave-vector, $\RO = \kwvZ \tuberadius$, for a warm dense tube (see \tref{tbl:cases}) are shown in \fref{fig:case:1:dispersion}. For this case the ordering of the characteristic speeds is as follows: $\vSE > \vSI > \vAE > \vAI$. In this figure, and in the following, we over-plot two cases, i) $\Beqphii / \Beqzi = 0.01$ and, ii) $\Beqphii / \Beqzi = 0.2$, which correspond to (practically) no twist and small twist, respectively. The reason for using a small, but non-zero twist for the case corresponding to the dispersion relation with zero twist which we have shown to be equivalent to the result of \cite{edwin1983wave}, is that the limits of the Kummer functions in \eref{eqn:kummer:limits:m:pos} and \eref{eqn:kummer:limits:m:neg} require an increasing number of terms as $a_i \rightarrow 0$ and their calculation becomes inefficient by direct summation. However, $\Beqphii / \Beqzi = 0.01$ is a good approximation to the case with zero azimuthal magnetic field component. Note that in the following we take the twist, namely $\Beqphi(r)$, to be continuous across the flux tube and thus set, $\TwistConstI = \TwistConstE$. The behavior of the fast sausage body waves (FSBW) is very similar for both the case with and without twist, and in extension it is very similar to the case with only internal twist studied by \cite{erdelyi2007linear}. It is worth noting that when internal and external twist is present, the different radial harmonics of the FSBW modes have two solutions, one dispersive and one approximately non-dispersive see \fref{fig:case:1:dispersion}. It is, however, unclear if the non-dispersive solution remains valid until the next radial harmonic. Nevertheless, it is clear that, in the neighborhood of the $\vSE$ \textit{singularity} we obtain two solutions with comparable phase speeds ($\vF$) which opens the possibility for beat phenomena and thus widens the possibility of detection of these waves since the beat frequency will be smaller than both waves that produce it. This behavior is not present when we consider twist only inside the flux tube. Otherwise, the overall behavior of the solutions is virtually identical to \cite{erdelyi2007linear}.

\begin{figure}
\includegraphics[width=\columnwidth]{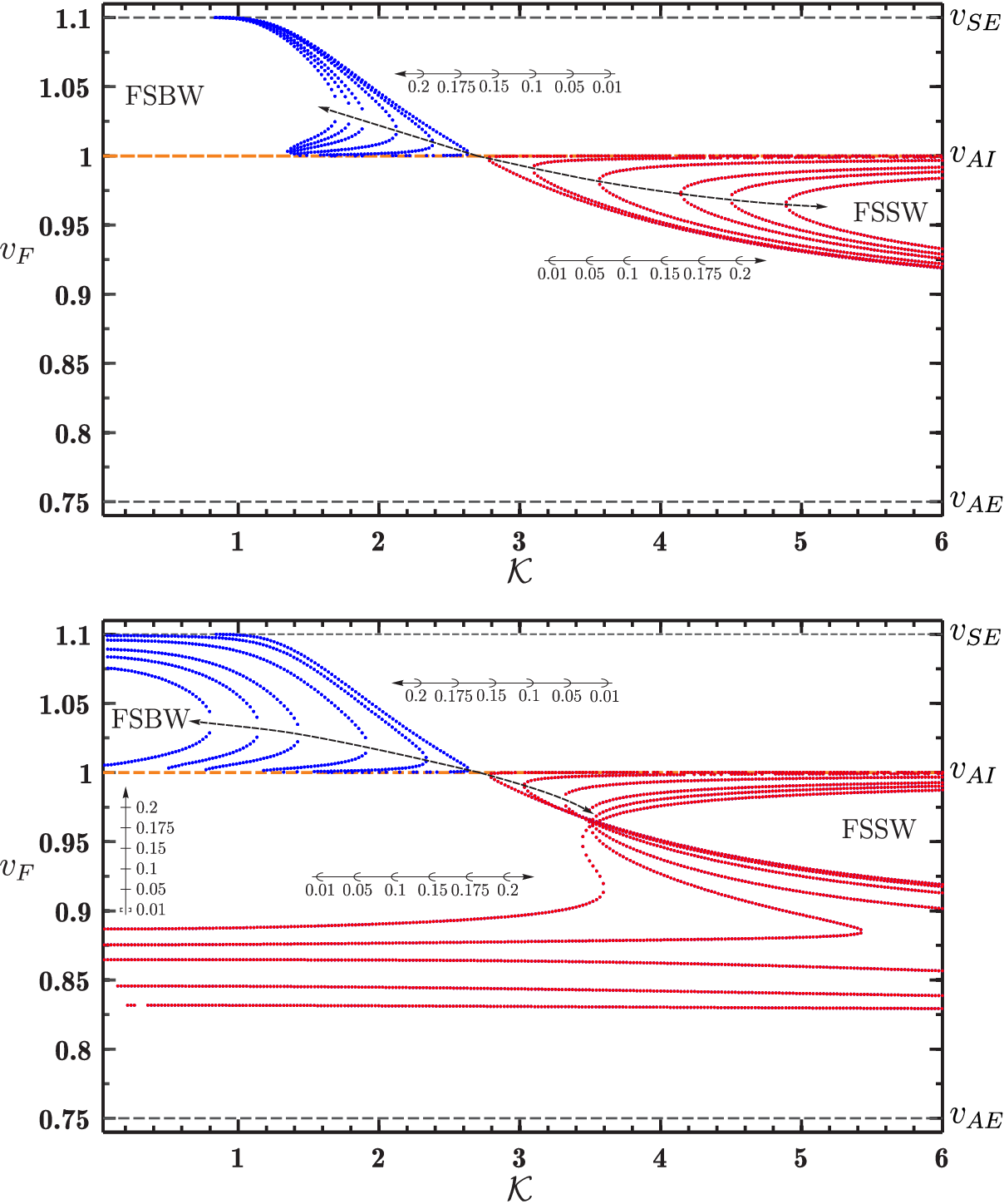}
\caption{Solutions of the dispersion \eref{eqn:dispersion:equation} for a cool evacuated tube embedded in a dense environment ($\beta_i \ll 1, \beta_e \gg 1$) with speed ordering $\vSE > \vAI > \vAE > \vSI$. The top figure corresponds to no external twist for $r>\tuberadius$, namely $\TwistConstE = 0$, while in the lower plot there is twist outside as well as inside the flux tube. Note that in both figures the solutions for $\Beqphii / \Beqzi = \{0.01, 0.05, 0.1, 0.15, 0.175, 0.2\}$ have been over-plotted to conserve space and illustrate the effect of increasing the magnetic twist. The axes inside the figures match the progression of twist for the specific regions, for instance, in the top plot the axis with the arrow to the left indicates that the first FSBW from the right corresponds to magnetic twist of $0.01$ the second to $0.05$ etc. Note that the vertical axis, for $\Beqphii / \Beqzi = 0.01$ has no non-dispersive FSSW (horizontal solutions marked in red) which is indicated by the empty parenthesis near the value $0.01$. The mean RMSE is $0.021$. \label{fig:case:2:dispersion}}
\end{figure}

\begin{figure}
\includegraphics[width=\columnwidth]{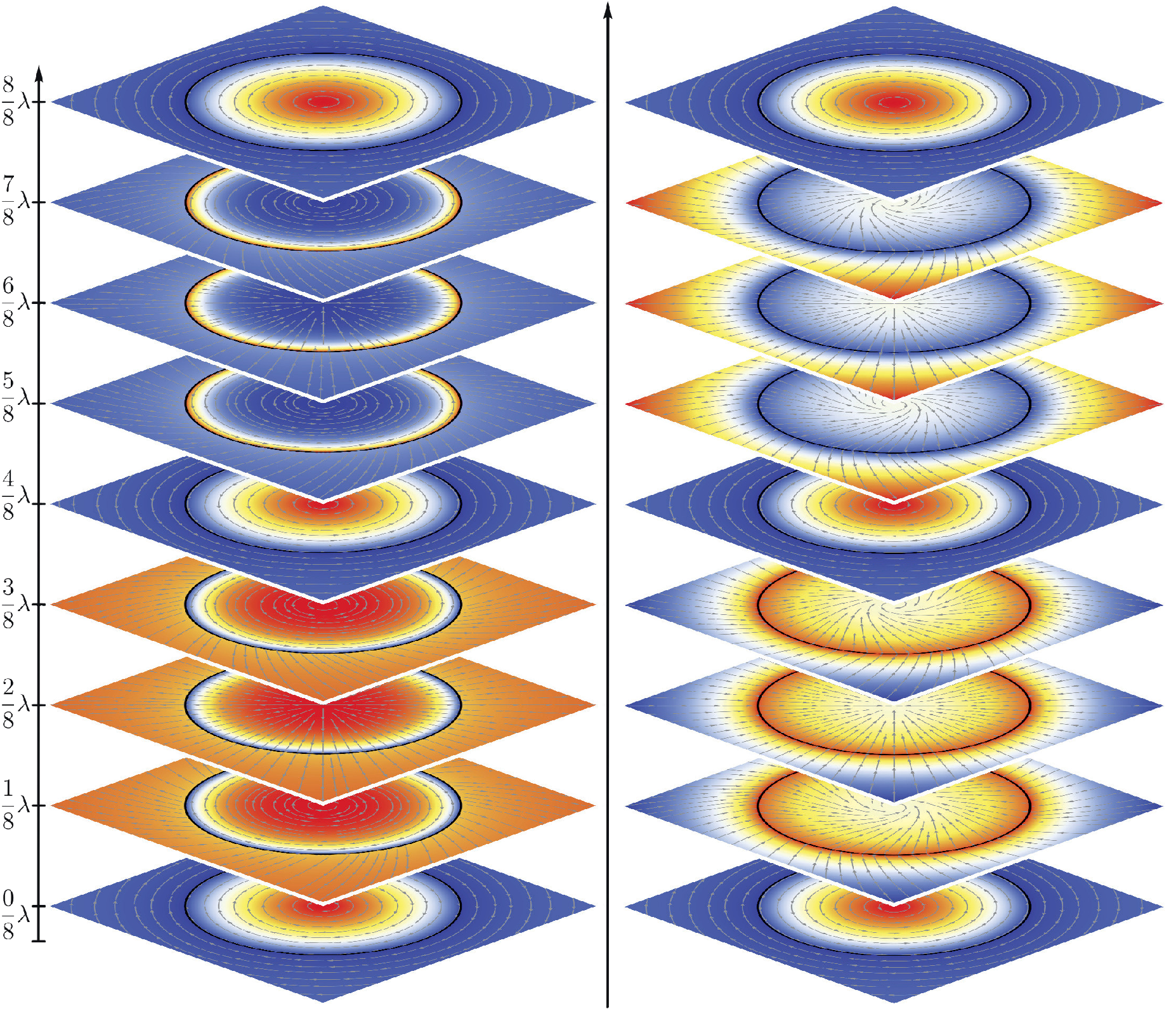}
  \caption{Plots of density perturbations superimposed on the background equilibrium density plots for the fast body and surface modes shown in \eref{fig:case:2:dispersion}. The gray lines represent velocity perturbation streamlines on the $xy$-plane. Notice that for visualization purposes the streamlines contain no information on the magnitude of the perturbation, only direction information is conveyed. In the density plots red and blue correspond to high and low density respectively. The slices are snapshots at $t=0$ at different positions for the wavelength $\lambda$ of the oscillation. Note that this does not imply that the wavelength of the two oscillations is the same, rather, it is a fraction of the corresponding wavelength. (\textbf{Left}) Fast body mode for with magnetic twist for $\RO = 0.3624$ and $\vF = 1.071$. (\textbf{Right}) Fast surface mode with magnetic twist for $\RO = 3.494$ and $\vF = 0.9058$. Notice that in both cases the azimuthal component of the velocity perturbation at $0/8 \lambda, 4/8 \lambda$ and $8/8 \lambda$ is is non-zero.\label{fig:case:2:2d:slices}}
\end{figure}

In \fref{fig:case:2:dispersion} we present the solutions for the second case in \tref{tbl:cases}. This scenario can occur when the internal plasma-$\beta$ is very low, $\beta_i \ll 1$, while the external plasma-$\beta$ is high, $\beta_e \gg 1$. At this point we would like to stress the fact that for a specific set of characteristic speeds the internal and external plasma-$\beta$ values are uniquely defined (see \axref{apx:speeds:ordering}). We focus here only in the region of solutions in $(\vAE,\vSE)$ since the infinite number of the slow sausage body waves (SSBW), present in the $(\vTI, \vSI)$ interval are minimally affected by the twist and thus are almost identical when compared with the corresponding case with no twist. We plot solutions for $\Beqphii / \Beqzi = \{0.01, 0.05, 0.1, 0.15, 0.175, 0.2\}$. The upper plot in \fref{fig:case:2:dispersion} represents the solutions only for internal twist. A feature for this set of solutions is that the FSBW which is transformed to the fast sausage surface (FSSW) for $\RO \sim 2.7$, for increasing twist the transition becomes discontinuous and an interval, in $\RO$, is created where there exist no solutions. For example, for $\Beqphii / \Beqzi = 0.05$, this interval extends for $\RO = (2.4,3.1)$ where no fast body waves exist. This interval becomes larger with increasing twist. However, this not the case in the presence of external twist, see lower figure in \fref{fig:case:2:dispersion}. The FSBW and FSSW appear to behave similarly, however, in all cases except for $\Beqphii / \Beqzi = 0.01$ there is a surface wave solution that is nearly non-dispersive for a wide range of $\RO$. Another feature is the \textit{s}-like set of surface wave solutions that are clearly visible for $\Beqphii / \Beqzi = 0.175$ and $\Beqphii / \Beqzi = 0.2$. Note, that this \textit{s}-like set is also present for the other cases however the cusp is encountered for larger values of $\RO$. This structure is quite interesting since in some interval of $\RO$ there exist $3$ simultaneous solutions while outside of this interval exists only one. This means that within that interval, for a broadband excitation, the power of the driver will be distributed to more than one solution thus reducing the individual power spectrum signatures of the individual waves. In essence this will result in a interval of solutions that are much more difficult to detect. Another interesting point in respect to this \textit{s}-like set of solutions is that it seems that a point may exist, for a certain value of $\Beqphii / \Beqzi$ and a single $\RO$ that there would be a continuum as the \textit{s}-shape becomes vertical (see \fref{fig:case:2:dispersion}). However, the existence or physical significance of this point is speculative since it does not appear to exist for small twist, namely the regime for which our approximation is valid. In \fref{fig:case:2:2d:slices} we illustrate a FSBW (left panel) and a SSBW (right panel). In contrast to the kink mode in magnetic flux tubes with weak twist that exhibit a polarization \citep{terradas2012transverse}, the sausage mode appears to be the a superposition of an Alfv\'{e}nic wave and a sausage wave leading by $\pi/2$.

\begin{figure}
\includegraphics[width=\columnwidth]{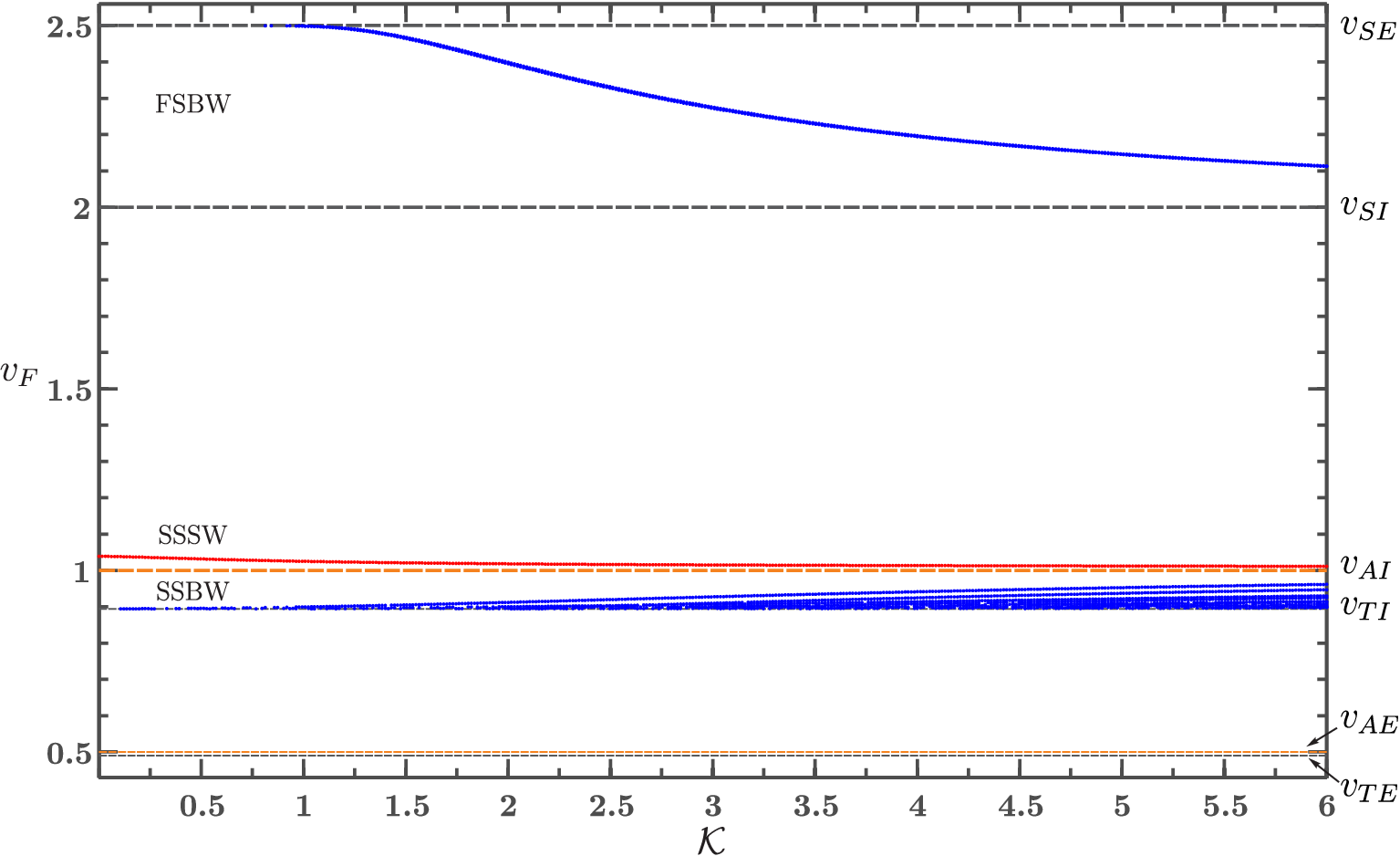}
\caption{Solutions of the dispersion \eref{eqn:dispersion:equation} for a weak cool tube embedded in a dense environment ($\beta_i >1, \beta_e \gg 1$) with speed ordering $\vSE > \vSI > \vAI > \vAE$. The mean RMSE is $0.0547$. \label{fig:case:4:dispersion}}
\end{figure}
The last plasma regime with high plasma-$\beta$ considered in this work has the following characteristic speed ordering: $\vSE > \vSI > \vAI > \vAE$. In \fref{fig:case:4:dispersion} we plot the solutions to \eref{eqn:dispersion:equation} for this case. The most notable feature, which seems to be consistent for alternative parameter sets corresponding to photospheric conditions, is that the magnetic twist appears to have only a small effect on the solutions to the dispersion equation. For example, we have also used: $\vAI > \vSE > \vSI > \vAE$ and there too (plot not shown as it is identical to the case with no twist) the deviation of the normalized phase speed was on the order of $0.5\%$ or less for magnetic twist up to $\Beqphii / \Beqzi = 0.2$.

\subsection{Low plasma-$\beta$ regime}\label{sec:low:plasma:beta}
\begin{figure}
\includegraphics[width=\columnwidth]{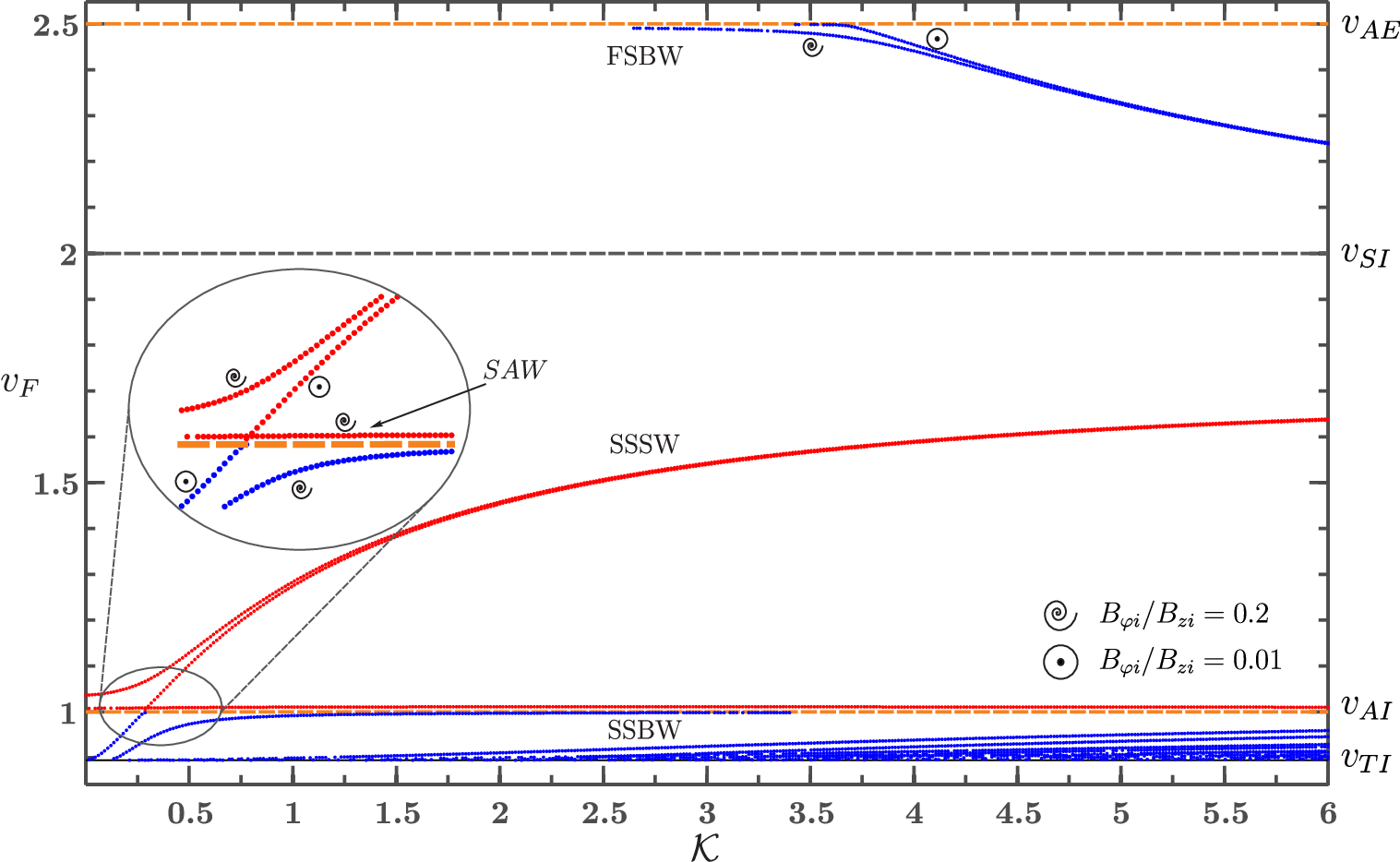}
\caption{Solutions of the dispersion \eref{eqn:dispersion:equation} for an intense warm tube embedded in a rarefied environment ($\beta_i \ll 1, \beta_e \ll 1$) with the following speed ordering, $\vAE > \vSI > \vAI > \vSE$. The mean RMSE is $0.0291$. \label{fig:case:6:dispersion}}
\end{figure}
In consultation with the results presented by \cite{vernazza1981structure} and \cite{gary2001plasma}, we expect the results presented in this section to be most relevant to conditions that are typical of the upper chromosphere the transition region and corona. The remaining two cases that we consider in this work are for low plasma-$\beta$ conditions (see \tref{tbl:cases}).

In \fref{fig:case:6:dispersion} we consider an intense warm flux tube for which the characteristic speeds ordering is the following: $\vAE > \vSI > \vAI  > \vSE$. This case was also considered by \cite{erdelyi2007linear} under the assumption that there is only internal magnetic twist and zero twist in the environment surrounding the flux tube. In that work the influence of twist was under a percent, however, when the external twist is also considered interesting behavior emerges. In this case, when there is zero twist, the first SSBW changes character to a slow sausage surface wave (SSSW) crossing $\vAI$ at approximately $\RO = 0.25$. When a small twist is introduced the first radial harmonic of the SSBW modes now becomes bounded by $\vAI$ and a SSSW mode appears. Also, a non-dispersive solution with a character similar to a surface wave emerges that closely follows $\vAI$. We have named this solution as surface-Alfv\'{e}n wave (SAW) in \fref{fig:case:6:dispersion} and we have expanded the plot to make it visible since it is extremely close to the internal Alfv\'{e}n speed. Interestingly the higher radial harmonics of the SSBW appear to be minimally affected when the magnetic twist is increased. Also, the correction to the phase velocity for the FSBW with magnetic twist appears to be small compared with the case of no magnetic twist. For the first radial harmonic this correction is of the order of $0.4\%$ while the correction is less than $0.1\%$ for higher radial harmonics. However, this does not mean that the FSBW for the case with magnetic twist is identical to the case without twist. This is because the azimuthal component of the velocity perturbation in the former case is non-zero altering the character of these waves significantly as compared with its counterpart in the case without magnetic twist.

\begin{figure}
\includegraphics[width=\columnwidth]{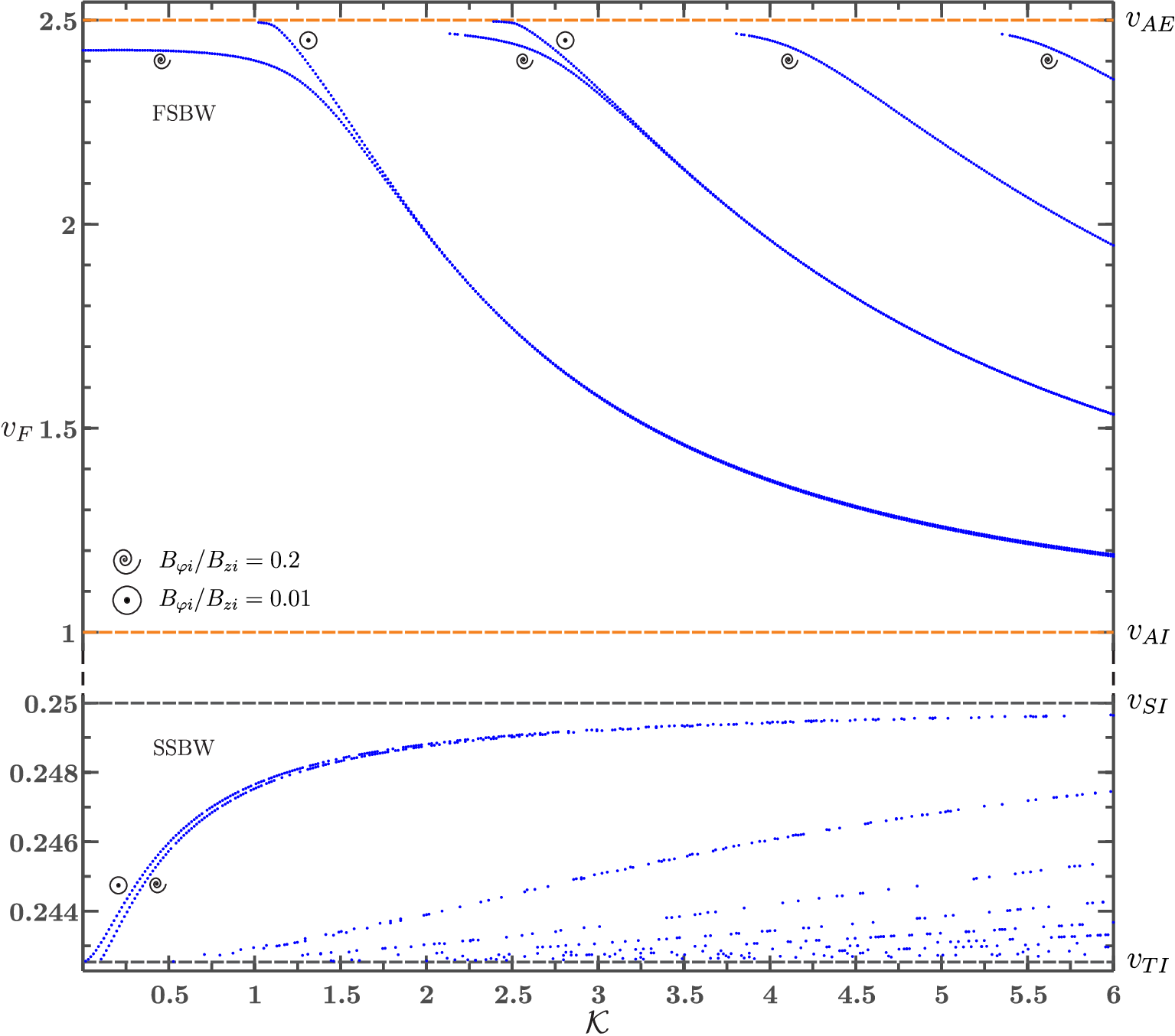}
\caption{Solutions of the dispersion equation \eref{eqn:dispersion:equation} for an intense cool tube embedded in a rarefied environment ($\beta_i \ll 1, \beta_e \ll 1$) with speed ordering $\vAE > \vAI > \vSE > \vSI$. The mean RMSE is $0.0175$. \label{fig:case:7:dispersion}}
\end{figure}
Lastly an intense cool tube is considered, i.e. $\vAE > \vAI > \vSE > \vSI$, which corresponds to conditions in the solar corona. The solutions to the dispersion equation (i.e. \eref{eqn:dispersion:equation}) are shown in \fref{fig:case:7:dispersion}. In this case, magnetic twist has more pronounced effect on the FSBWs, while the SSBW are virtually unaffected. In the long wavelength limit, $\RO \ll 1$, the FSBWs become non-dispersive while for short wavelength limit, $\RO \gg 1$, the solutions are identical to the case of a straight magnetic flux tube with vertical magnetic field only. It is important to note that, although the effect of magnetic twist appears to be subtle in this case, it has an significant difference compared with the case with no magnetic twist, e.g. \cite{edwin1983wave}, as well as the case considering only internal magnetic twist, e.g. \cite{erdelyi2007linear}. In both of these cases the sausage mode becomes leaky as $\RO \rightarrow 0$. This however, is not the case when both internal and external twist are considered. Instead, as the magnetic twist increases so does the cutoff of the trapped fast sausage waves toward longer wavelengths. For example, for the particular characteristic speeds ordering considered in \fref{fig:case:7:dispersion}, the first FSBW ceases to have a cutoff wavelength when $\Beqphii / \Beqzi > 0.05$, approximately. Therefore, the FSBW for a twisted magnetic cylinder above a certain threshold remains trapped for all wavelengths. A consequence of this is that FSBWs remain in the Alfv\'{e}n continuum and therefore may be resonantly damped, see for example \cite{sakurai1991resonantI}. This means that the sausage mode cannot be ruled out as a source of energy to the corona.

\section{Discussion}\label{sec:discussion}
Although the model we present in this work for a magnetic flux tube with internal and external twist is relatively advanced in comparison to recent theoretical models, it contains a number of simplifications and therefore we would be remiss not to discuss the potential caveats when used to interpret observations. Observations suggest that the cross-section of magnetic flux tubes is not circular. Although there are no theoretical studies of magnetic flux tubes with completely irregular cross-section, some steps towards this direction have been taken by considering flux tubes with elliptical cross-section, see for example \cite{ruderman2003resonant} and \cite{erdelyi2009magnetohydrodynamic}. The results for the sausage mode presented in \cite{erdelyi2009magnetohydrodynamic} show that in comparison with the model of \cite{edwin1983wave} (circular cross-section) the ellipticity of the cross-section tends to increase the phase speed of the sausage mode for photospheric conditions by approximately $5\%$ in the short wavelength limit, and, is negligible in the long wavelength limit. Conversely, in coronal conditions for increasing ellipticity the phase speed increase is more pronounced for a wide range of wavelengths and is shown to be as much as $20\%$ higher of the predicted phase speed by the model with circular cross-section. This effect is quite important since, for sufficiently large ellipticity, it could counteract the effect that magnetic twist has on the cutoff frequency for the fast sausage body modes seen in \fref{fig:case:7:dispersion}. Namely, as can be seen in \fref{fig:case:7:dispersion}, the fast sausage mode remains trapped in the long wavelength limit, however, should the phase speed be increased, then a cutoff frequency for the fast sausage body modes may be reinstated.

Furthermore, although we have studied propagating waves in this work, the study of standing modes for $\kwvTh = 0$ is trivially extended. Namely, if the magnetic flux tube is line-tied on both footpoints the longitudinal wavevector will be quantized according to $\kwvZ = n \pi / L$, where $n$ is an integer and $L$ is the length of the magnetic flux tube. If the flux tube is assumed to be line tied on one end and open on the other, then no quantization takes place and there can be propagating and standing waves for all $\kwvZ$. Here it should be mentioned that the effect of the magnetic flux tube curvature is of the order of $(\tuberadius / L)^2$ and therefore has a small effect on the eigenfrequencies of magnetic flux tubes in the solar atmosphere \citep{van2004effect,van2009effect}.

Other effects that can alter the eigenfrequencies predicted using the model in this work are, density stratification, flux tube expansion and resonance phenomena due to neighboring magnetic flux tubes, see \cite{ruderman2009transverse} for a more in depth discussion. Of course, more complicated magnetic field topologies can have other unforeseen effects. This can be seen in magneto-convection simulations, e.g. \cite{wedemeyer2012magnetic}, \cite{shelyag2013alfv}, \cite{trampedach2014improvements} as well as in simulations with predefined background magnetic fields, see \cite{bogdan2003waves}, \cite{vigeesh2012three}, \cite{fedun2011mhd}. However, the interpretation of the results from such simulations is a major challenge which is only increased by considering that the initial conditions, which are mostly unknown, play a very important role in their subsequent evolution.

\section{Conclusions}\label{sec:conclusions}
In the presence of weak twist the sausage mode has mixed properties since it is unavoidably coupled to the axisymmetric Alfv\'{e}n wave. This apparent from the solutions, see for example \axref{apx:perturbed:quantities} where the azimuthal velocity perturbation component is nonzero and is also supported by the results presented in \sref{sec:governing:equations}. The implications of this on the character of surface and body waves are seen clearly in \fref{fig:case:2:2d:slices}, where the relative magnitude of the radial and azimuthal components of the velocity perturbation alternate periodically and waves tend to exhibit Alfv\'{e}nic character the closer their phase velocity is to one of the Alfv\'{e}n speeds. The reason for this behaviour has been explained in \sref{sec:governing:equations}.

Observations of Alfv\'{e}n waves rely on the apparent absence of intensity (i.e. density) perturbations in conjunction with torsional motion observed by alternating Doppler shifts, see for example \citep{jess2009alfven}. The results of this work suggest that there exists at least one more alternative interpretation for waves with the aforementioned characteristics. Namely, the observed waves by \cite{jess2009alfven} could potentially be surface sausage waves (see right panel of \fref{fig:case:2:2d:slices}), since the localized character of the density perturbation also implies localized intensity perturbations that can be well below the instrument resolution. Furthermore, given the presence of torsional motion (see right panel of \fref{fig:case:2:2d:slices}) the sausage mode will have a Doppler signature similar to that of an Alfv\'{e}n wave. The Doppler signature in combination with the fact that surface waves can have a phase velocity very close to the Alfv\'{e}n speed (see SAW in \fref{fig:case:6:dispersion}) suggests that \cite{jess2009alfven} potentially observed a sausage mode in the presence of magnetic twist. This line of reasoning is further supported by the evidence in \cite{wedemeyer2012magnetic} and \cite{morton2013evidence}, where the authors show that vortical motions are ubiquitous in the photosphere. However, the excitation of the decoupled Alfv\'{e}n wave requires vortical motion that is divergence free, see for example \cite{ruderman1997direct}, while the vortical motions observed in \cite{morton2013evidence} are not free of divergence. In our view these vortical motions could be a natural mechanism for the excitation of the axisymmetric modes studied in this work.

In this work we considered the effect of internal and external magnetic twist on a straight flux tube for the sausage mode. It was shown that magnetic twist naturally couples axisymmetric Alfv\'{e}n waves with sausage waves. Some of the main results of this coupling are:
\begin{itemize}
\item Sausage waves can exhibit Doppler signatures similar to these expected to be observed for Alfv\'{e}n waves.
\item The phase difference between the radial and torsional velocity perturbations are $\pi/2$, which means that both effects can be simultaneously observed.
\item Excitation of these modes can be accomplished with a larger variety of drivers compared to the \textit{pure} sausage and axisymmetric Alfv\'{e}n waves. Therefore, we speculate that these waves should be more likely to be observed compared with their decoupled counterparts.
\item For coronal conditions the fast sausage body waves remain trapped for all wavelengths when the magnetic twist strength surpasses a certain threshold. This appears to be characteristic of magnetic twist and could potentially be used to identify the strength of the magnetic twist.
\end{itemize}
These findings suggest that axisymmetric modes with magnetic twist can be easily mistaken for pure Alfv\'{e}n waves.

%
\appendix
\section{Dimensionless Dispersion Equation}\label{apx:non:dimensional:dispersion}
For completeness we give here the dimensionless version of the dispersion equation \eref{eqn:dispersion:equation}. The following equation is now a function of $\vF$ and $\RO$, instead of $\omega$ and $\kwvZ$. One of the benefits of solving \eref{eqn:nondim:dispersion:equation} instead of \eref{eqn:dispersion:equation} directly is that the former is, usually, numerically more stable. Another benefit is that the study of different plasma conditions is made simpler since it is straightforward to alter the ordering of the characteristic speeds ($\vSi,\, \vAi$ etc).

\begin{dmath}\label{eqn:nondim:dispersion:equation}
\frac{\rhoe}{\rhoi}\frac{\RO (\vF^2 - \vAE^2)}{\kwvROUT^2} \frac{K_{\NU-1}(\kwvROUT \RO) }{K_{\NU}(\kwvROUT \RO )} = \left[ \frac{\vAphI^2}{\kwvRIN^2}(1+\NI)^2 - \frac{\rhoe}{\rhoi}\frac{\vAphE^2}{\kwvROUT^2}(1+\NE^2) \right] +\frac{\rhoe}{\rhoi}\frac{(1-\NU)(\vF^2 - \vAE^2)}{M_E^2} - 2\frac{(\vF^2 -1)}{\kwvRIN^2}\frac{\KummerM{\kummerA}{\kummerB-1}{\kummerVar}}{\KummerM{\kummerA}{\kummerB}{\kummerVar}},
\end{dmath}
where,
\begin{align*}
\kummerA = 1 + \frac{ \kwvRIN^2\left[ \RO^2(\vF^2-1)^2 - 4 \vAphI^2 \right] }{ 8 \vAphI^2 \NI (\vF^2 - 1) }, && \kummerB = 2, && \kummerVar = 2 \frac{\vAphI^2 \NI}{(\vF^2 - 1)},
\end{align*}
\begin{align*}
\vph = \frac{\omega}{\kwvZ}, && \vF = \frac{\vph}{\vAi}, && \vSI = \frac{\vSi}{\vAi}, && \vSE = \frac{\vSe}{\vAi}, \\
\vAI = 1, && \vAE = \frac{\vAe}{\vAi},  && \Ni^2 = \kwvZ^2 \NI^2, && \Ne^2 = \kwvZ^2 \NE^2,\\
\kwvRin^2 = \kwvZ^2 \kwvRIN^2, && \kwvRout^2 = \kwvZ^2 \kwvROUT^2 && \vTI = \frac{\vTi}{\vAi}, && \vTE = \frac{\vTe}{\vAi},\\
\RO = \kwvZ \tuberadius, && \vAphI = \frac{\vAphi}{\vAi}, && \vAphE = \frac{\vAphe}{\vAi},&& 
\end{align*}
and
\begin{align*}
\NI^2 = \frac{\vF^4}{\vF^2 + \vSI^2(\vF^2 -1)}, && \NE^2 = \frac{\vF^4}{\vF^2 \vAE^2 + \vSE^2(\vF^2 - 1)}, \\
\kwvRIN^2 = \frac{(\vSI^2 - \vF^2)(1 - \vF^2)}{(1 + \vSI^2)(\vTI^2 - \vF^2)}, && \kwvROUT^2 = \frac{(\vSE^2 - \vF^2)(\vAE^2 - \vF^2)}{(\vAE^2 + \vSE^2)(\vTE^2 - \vF^2)},
\end{align*}
\begin{align*}
\vTi^2 = \frac{\vAi^2 \vSi^2}{\vAi^2 + \vSi^2}, && \vTI^2 = \frac{\vSI^2}{1 + \vSI^2}, && \vTe^2 = \frac{\vAe^2 \vSe^2}{\vAe^2 + \vSe^2},\\
\vTE^2 = \frac{\vAE^2 \vSE^2}{\vAE^2 + \vSE^2}, && &&
\end{align*}
\begin{dmath*}
\nu^2 = 1 + 2 \frac{\vAphE^2}{(\vF^2 - \vAE^2)^2}\left[ 2 \vAphI^2 \NE^2 + (\vAE^2(3 \NE^2 - 1) - \vF^2(\NE^2 +1)) \right].
\end{dmath*}
Also the plasma-$\beta$ inside and outside the flux-tube can be calculated using: $\beta_i = (2/\gamma) \vSI^2$ and $\beta_e = (2 \rhoe B_{z_i}^2 /\gamma \rhoi B_{z_e}^2) \vSE^2$ respectively.

\section{Estimation of the Root Mean Square Error}\label{apx:rmse}
We argue that the exact solution for constant twist outside the magnetic flux tube is a good approximation to the solution corresponding to the case where the twist is $\propto 1/r$. However, as we state in the text, we only obtain the zeroth-order term in the perturbation series which corresponds to constant twist. To justify this statement we estimate the root mean squared error (RMSE) also referred to as standard error, defined as follows,
\begin{align*}
RMSE = \underset{L\rightarrow \infty}{\lim}\left( \frac{1}{L-1} \int_{r_a}^L \left( \hat{\Xirep}(r) - \Xirep(r) \right)^{2} \, dr \right)^{\frac{1}{2}}.
\end{align*}
In this context, $\hat{\Xirep}(r)$ is the solution to the case with constant magnetic twist, i.e. $\mee = 0$ in \eref{eqn:xirode:outside}, while $\Xirep(r)$ is a numerical solution to \eref{eqn:xirode:outside} with $\mee = 1$, namely magnetic twist proportional to $1/r$. The RMSE is expected to vary for different parameters, i.e. $\RO,\vF$ and $\Beqphii / \Beqzi$, and for this reason we discretize $\RO$ and $\vF$ using a $100\times100$ grid and also use the following value for $\Beqphii / \Beqzi = 0.2$, since for all values of $\Beqphii / \Beqzi < 0.2$ the RMSE is consistently smaller. Subsequently we average the resulting root mean square errors which we then quote in the corresponding figure caption. Note, that $\hat{\Xirep}(r)$ and $\Xirep(r)$ are normalized, therefore a value for the mean RMSE of, e.g. $0.01$, means that the standard error is $1\%$ on average, when comparing $\hat{\Xirep}(r)$ with $\Xirep(r)$.

\section{Characteristic Speeds Ordering Considerations}\label{apx:speeds:ordering}
The ordering of characteristic speeds depends on $6$ variables: $\Beqzi$, $\Beqze$, $T_i$, $T_e$, $n_i$, $n_e$, where $n_i$ and $n_e$ are the number densities inside and outside the flux tube respectively. Starting from and assuming the magnetic twist is small,
\begin{align*}
\vAi = \frac{\Beqzi}{\sqrt{\mu_0 \rhoi}}, && \vAi = \frac{\Beqzi}{\sqrt{\mu_0 \rhoi}}, && \vSi = \sqrt{ \gamma \frac{p_i}{\rhoi} }, && \vSe = \sqrt{ \gamma \frac{p_e}{\rhoe} }, \\
\beta_i = \frac{2 \mu_0 p_i}{\Beqzi^2}, && \beta_e = \frac{2 \mu_0 p_e}{\Beqze^2}, && \rho = n m_p, && p = n k_B T.
\end{align*}
Taking logs of the speeds and using the following definitions:
\begin{align*}
\vAi^{\star} = \ln(\vAi) + \frac{1}{2}\ln(\mu_0 m_p), &&
\vAe^{\star} = \ln(\vAe) + \frac{1}{2}\ln(\mu_0 m_p), \\
\vSi^{\star} = \ln(\vSi) - \frac{1}{2}\ln\left( \gamma \frac{k_B}{m_p} \right), &&
\vSe^{\star} = \ln(\vSe) - \frac{1}{2}\ln\left( \gamma \frac{k_B}{m_p} \right), \\
b_i^{\star} = \frac{1}{2}\left( \ln(\beta_i) - \ln(2 k_B \mu_0) \right), &&
b_e^{\star} = \frac{1}{2}\left( \ln(\beta_e) - \ln(2 k_B \mu_0) \right), \\
\end{align*}
and
\begin{align*}
\Beqzi^{\star} = \ln(\Beqzi), && \Beqze^{\star} = \ln(\Beqze), && T_i^{\star} = (1/2)\ln(T_i), \\
T_e^{\star} = (1/2)\ln(T_e), && n_i^{\star} = (1/2)\ln(n_i), && n_e^{\star} = (1/2)\ln(n_e),
\end{align*}
the speeds and plasma-$\beta$ parameters can be written as follows,
\begin{dmath*}
\left(
\begin{array}{cccccc}
 1 & 0 & 0 & 0 & -1 & 0 \\
 0 & 1 & 0 & 0 & 0 & -1 \\
 0 & 0 & 1 & 0 & 0 & 0 \\
 0 & 0 & 0 & 1 & 0 & 0 \\
 -1 & 0 & 1 & 0 & 1 & 0 \\
 0 & -1 & 0 & 1 & 0 & 1 \\
\end{array}
\right) \left(
\begin{array}{c}
 \Beqzi^{\star} \\
 \Beqze^{\star} \\
 T_i^{\star} \\
 T_e^{\star} \\
 n_i^{\star} \\
 n_e^{\star} \\
\end{array}
\right)=\left(
\begin{array}{c}
 \vAi^{\star} \\
 \vAe^{\star} \\
 \vSi^{\star} \\
 \vSe^{\star} \\
 b_i^{\star} \\
 b_e^{\star} \\
\end{array}
\right).
\end{dmath*}
Now notice that the above matrix is rank $4$ which means that the dimension of the null-space is $2$, with basis vectors: $y_1 = (1,0,0,0,1,0)$, and, $y_2 = (0,1,0,0,0,1)$. This means in practice that for a given set of parameters resulting in a specific speed ordering $\beta_i$ and $\beta_e$ are uniquely defined but there is a $2$ dimensional subspace involving $\Beqzi^{\star}$, $\Beqze^{\star}$, $n_i^{\star}$, $n_e^{\star}$, that is, all linear combinations of $y_1$ and $y_2$. Also notice that the sound speeds, $\vSi^{\star}$ and $\vSe^{\star}$, depend only on the internal and external temperature, $T_i^{\star}$ and $T_e^{\star}$ respectively. Additionally the null-space of the matrix (see the basis vectors $y_1$ and $y_2$) suggests that the densities, $n_i^{\star}$ and $n_e^{\star}$, are secondary variables to the magnetic field strength, $\Beqzi^{\star}$ and $\Beqze^{\star}$.

\section{Perturbed quantities in terms of $\Xirp$ and $\pTp$}\label{apx:perturbed:quantities}
Given, $\Xirp$ and $\pTp$ in \eref{eqn:xir:inside:solution},\eref{eqn:pti:inside:solution} or \eref{eqn:xir:outside:solution},\eref{eqn:pte:outside:solution} the remaining perturbed quantities for the sausage mode ($\kwvTh=0$) are \citep{erdelyi2010magneto},
\begin{dgroup*}
\begin{dmath*}
\Xirip(\kummerVar) = A_{i 1} \frac{\kummerVar^{1/2}}{E^{1/4}} e^{-\kummerVar/2} \KummerM{\kummerA}{\kummerB}{\kummerVar},
\end{dmath*}
\begin{dmath*}
\pTip(\kummerVar) = A_{i 1} \frac{k_a D_i}{\Ni^2 - \kwvZ^2} e^{-\kummerVar/2}\left[ \frac{\Ni+\kwvZ}{\kwvZ} \kummerVar \KummerM{\kummerA}{\kummerB}{\kummerVar} - 2 \KummerM{\kummerA}{\kummerB-1}{\kummerVar} \right],
\end{dmath*}
\begin{dmath*}
\Xirep(r) = A_{e 1}K_{\nu}\left(\kwvRout r\right),
\end{dmath*}
\begin{dmath*}
\pTep =
A_{e 1} \left(\frac{ \mu_0 (1-\nu) D_e - 2 \tuberadius^{2} \TwistConstE^2 n_e^2 }{\mu_0 r (\kwvZ^2 - \Ne^2)} K_{\nu}(\kwvRout r) - \frac{ D_e }{\kwvRout} K_{\nu-1}(\kwvRout r) \right),
\end{dmath*}
\begin{dmath*}
\Xiphip = \frac{i \kwvZ}{\rho (\omega^2 - \omega_A^2)} \frac{\Beqz \Beqphi}{\mu_0 \rho} \left[ \frac{n^2}{\omega^2} \left( \frac{2 \Beqphi^2}{\mu_0 r} \Xirp - \pTp \right) - 2 \frac{\rho}{r}\Xirp \right],
\end{dmath*}
\begin{dmath*}
\Xizp = \frac{i \kwvZ}{\rho \omega^2 (\omega^2 - \omega_A^2)} \left[ (\omega^2 -n^2 \vA^2)\pTp + 2\frac{ \Beqphi^2}{\mu_0 r} \vA^2 n^2 \Xirp  \right],
\end{dmath*}
\begin{dmath*}
\Brp = i \kwvZ \Beqz \Xirp,
\end{dmath*}
\begin{dmath*}
\Bphip = \frac{\kwvZ^2 \Beqphi}{D}\left( \pTp + 2 \frac{\Beqz^2}{\mu_0 r}\Xirp \right) - \D{}{r}(\Beqphi \Xirp),
\end{dmath*}
\begin{dmath*}
\Bzp = - \frac{1}{r} \D{}{r}(r \Beqz \Xirp).
\end{dmath*}
\end{dgroup*}
%
\section{Acknowledgments}
I.G. would like to acknowledge the Faculty of Science of the University of Sheffield for the SHINE studentship and M. Ruderman, T. Van Doorsselaere and M. Goossens for the useful discussions on this paper. V.F., G.V. and R.E. would like to acknowledge STFC for financial support. R.E. is also thankful to the NSF, Hungary (OTKA, Ref. No. K83133). G.V. would like to acknowledge the Leverhulme Trust (UK) for the support he has received.

%
%
\bibliographystyle{aa-note}

\end{document}